# HexRAN: A Programmable Approach to Open RAN Base Station System Design

Ahan Kak, *Member, IEEE*, Van-Quan Pham, *Member, IEEE*,
Huu-Trung Thieu, *Member, IEEE*, and Nakjung Choi, *Senior Member, IEEE*

***Abstract*—In recent years, the radio access network (RAN) domain has witnessed a sea change with increasing levels of virtualization and softwarization driven by emerging paradigms such as the Open RAN (O-RAN) movement. However, the fundamental building block of the cellular network, i.e., the base station, remains unchanged and ill-equipped to handle this architectural evolution. In particular, with reference to existing base station architectures, there exists a general lack of programmability and composability along with a protocol stack that grapples with diverging and often conflicting requirements set forth by 3GPP and O-RAN. Recognizing the need for an "O-RAN-native" approach to base station design, this paper introduces HexRAN–a novel base station architecture characterized by key features relating to RAN disaggregation and composability, 3GPP and O-RAN protocol integration and programmability, robust controller interactions, and customizable RAN slicing. Furthermore, the paper also includes a concrete systems-level prototype and comprehensive experimental evaluation of HexRAN on an over-the-air testbed, showcasing the scalability and performance benefits associated with the proposed architecture.**

## I. Introduction

CELLULAR networks have been increasingly characterized by the presence of general-purpose computing hardware and the softwarization of network functions (NFs). As networks evolve from 5G to 5G-Advanced and then 6G, these trends will continue to play a critical role in the design, management, and operation of radio access (RAN) and core networks [1], [2]. At the same time, these networks are also expected to serve a burgeoning set of application scenarios ranging from the more well-known enhanced mobile broadband (eMBB) to the upcoming immersive extended reality (XR), along with several automation-related use cases, each presenting a diverse set of service level agreement (SLA) requirements, necessitating the need for robust network programmability, closed-loop automation and control, and the provisioning of multiple network slices.

Recognizing the growing complexity of cellular networks, both academia [3]–[5] as well as the industry [6] have been engaged in efforts to redesign access networks based on the O-RAN concept [7]. Building upon 3GPP's NG-RAN architecture which provides the Central Unit-Control Plane (CU-CP), Central Unit-User Plane (CU-UP), and Distributed Unit (DU) base station components, O-RAN further disaggregates the DU into an O-RAN DU (O-DU) and an O-RAN RU (O-RU). Furthermore, through the introduction of a near-real time RAN Intelligent Controller (near-RT RIC) with supporting control applications called xApps and a non-real time RAN Intelligent Controller (non-RT RIC) with rApps, O-RAN has also given rise to the possibility of introducing novel network control functionalities and assurance frameworks. We note that this paper requires familiarity with the 3GPP and O-RAN architectures, systems, and interfaces. Interested readers may refer to §II-A for additional information.

However, while the RAN ecosystem has been evolving to support network programmability, the base station architecture is still primarily centered around 3GPP protocols and layers such as the MAC (Medium Access Control), PDCP (Packet Data Convergence Protocol), and SDAP (Service Data Adaptation Protocol). The current state-of-the art thus grapples with a fundamental dichotomy wherein O-RAN seeks to achieve programmable control and telemetry, but cellular base stations are not designed for such fine-grained interactions with other external entities. For example, configuration updates often require a full base station reset, thereby impacting network uptime. Furthermore, O-RAN brings with it a host of new specifications that seek to further complicate the already complex world of cellular standardization leading to network performance optimization (NPO) challenges. To that end, in order to leverage the full potential of O-RAN, it is imperative that base stations incorporate "O-RAN-native" design principles while still retaining the proven reliability of the 3GPP protocol stack. Additionally, this seamless blend of 3GPP and O-RAN philosophies should introduce minimal overhead while enabling streamlined NPO.

With a view to addressing the aforementioned key requirements, through this work, we introduce HexRAN, a programmable approach to O-RAN base station design, with several novel features relating to RAN disaggregation and composability, 3GPP and O-RAN protocol integration and programmability, robust RAN-RIC communications, and customizable network slicing as detailed next. More specifically, through HexRAN, we provide the following contributions.

- **Novel Paradigms for RAN Disaggregation and Composability:** Fully disaggregated by design, HexRAN introduces the possibility of dynamic service-specific compositions of different RAN functions to deliver tailored application-centric performance (§IV).
- **Programmable Mediation Layer for Streamlined Network Control and Assurance:** Recognizing the dichotomy between the 3GPP and O-RAN ecosystems, HexRAN introduces a Programmable Mediation Layer that provides abstractions for and interacts with the 3GPP

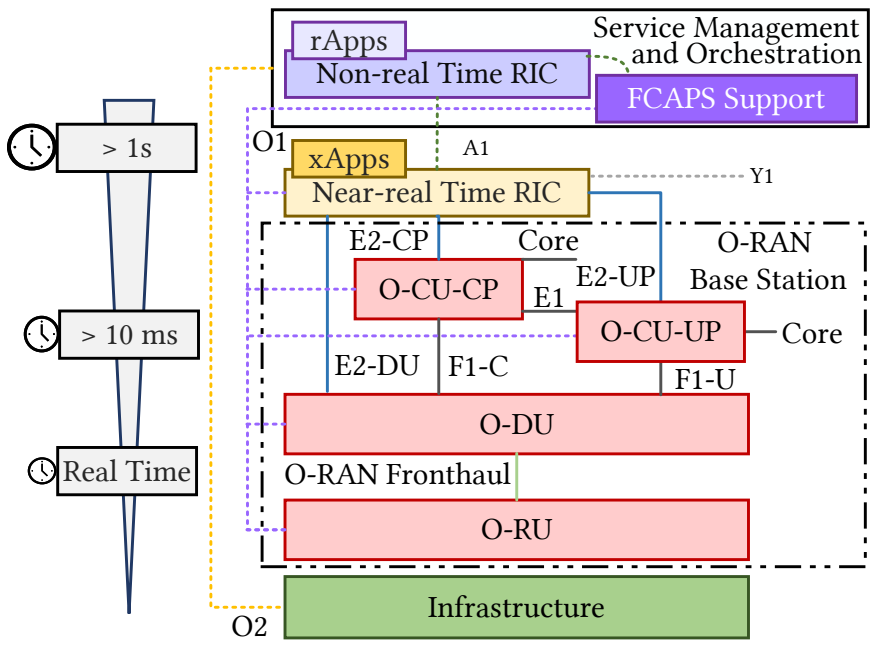

Fig. 1: The O-RAN reference architecture [7].

protocol stack to assist with system configuration and telemetry operations (§IV).

- **HexRAN Agent for Robust Extensibility:** With a focus on seamless RAN-RIC interactions, HexRAN introduces a highly extensible termination point for the E2 and O1 interfaces, called the HexRAN Agent. The HexRAN Agent is characterized by a modular high-performance architecture and plugin-based approach to service integration (§IV).
- **Vertically Integrated RAN Programmability:** The vertically integrated RAN programmability enabled by HexRAN is showcased through a RAN slicing use case that leverages and builds upon the aforementioned novel features with an emphasis on slice customizability and use case-driven operations (§V).
- **Concrete Systems-level Prototype:** In the interest of demonstrating its efficacy and practical applicability, a prototype of HexRAN is implemented and deployed on a large-scale over-the-air testbed for an extensive system evaluation and comparison with the state-of-the-art (§VI).

The remainder of this paper is organized as follows. In Section II, we present the relevant background concerning O-RAN as well as related work within the domain of base station programmability. Section III outlines the key requirements for achieving programmable base station design along with the associated challenges that must be overcome. Next, in Section IV, we introduce the HexRAN architecture with a view to addressing the aforementioned requirements, with Section V showcasing the RAN programmablility enabled by HexRAN through a RAN slicing use case. Section VI provides a comprehensive evaluation of the proposed architecture, and, Section VII finally concludes the paper.

## II. Background and Related Work

### A. O-RAN Architecture Overview

O-RAN introduces the core tenets of disaggregation and programmability to the RAN. As illustrated in Fig. 1, the O-RAN architecture consists of four broad categories of components– the physical infrastructure, on top of which are deployed the RAN elements, the near-RT RIC, and the Service Management and Orchestration (SMO) Framework [7]. The RAN can be deployed in either a monolithic or disaggregated configuration, with the latter including the CU-CP, the CU-UP, the DU, and the RU. The E1 interface links the CU-CP and CU-UP, which in turn communicate with the DU over the F1-C and F1-U interfaces respectively. Physical layer processing is split between the DU and RU which are linked together by the O-RAN Open Fronthaul interface (based on the 7-2x functional split [8]). RAN programmability is achieved through the introduction of two external network controllers, i.e., the near-RT and non-RT RICs.

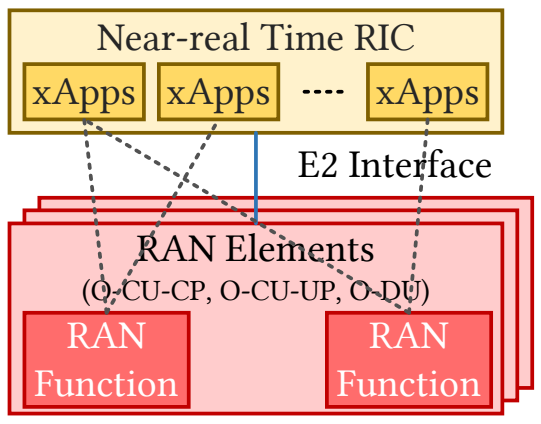

Fig. 2: Relationship between xApps and RAN functions [13].

More specifically, the RAN elements interact with the near-RT RIC over the E2 interface using the E2AP protocol, allowing for near-RT (i.e., > 10 ms) control of and telemetry from the underlying RAN elements through the xApps. The configuration and reporting features exposed by a RAN element depend upon the RAN functions supported by that element, characterized by a many-to-many mapping between RAN functions and xApps as shown in Fig. 2. Moreover, while the E2 interface and its associated procedures provide a generic RAN-RIC interaction framework, the constituent event triggers, telemetry metrics, and control parameters are defined in RAN function-specific Service Models (SMs). These function-specific E2SMs are specifications in their own right, describing the different RAN parameters that can be controlled and RAN statistics that can be reported back to the RIC. The current O-RAN specification includes four such SMs– Key Performance Measurement (KPM) [9], Network Interface (NI) [10], RAN Control (RC) [11], and Cell Configuration and Control (CCC) [12].

Finally, the Service Management and Orchestration (SMO) framework is responsible for providing fault, configuration, accounting, performance, security (FCAPS) support, in addition to incorporating the non-RT RIC that oversees the control and optimization of RAN elements on a non-real time, i.e., > 1 s, basis. Such control is enabled through applications called rApps that interact with the RAN elemenents over the NETCONF-based O1 interface. Interactions over O1 are codified in terms of YANG models representing the DU [14], and the CU-CP and CU-UP [15].

### B. Programmable Base Station Design

The introduction of the O-RAN architecture has resulted in a plethora of works that seek to optimize RAN performance ranging from ML-driven approaches to literature that seeks to address fundamental systems-level challenges. In particular, concerning the former, notable prior art includes ML-based xApps for near-RT RAN control [16], automated radio resource management for RAN slices [17], and solutions for mitigating jamming attacks [18]. On the other hand, concerning the latter we note the presence of solutions such as FlexSlice [19], ProSLICE [20], RAN Engine [21], all of which seek to achieve programmable RAN orchestration. More specifically, the authors in [19] have proposed enhancements to the O-RAN RAN slicing architecture with a view to reducing RAN-RIC control latency, similarly, the work presented in [20] provides abstractions for slice management and control through O-RAN. On the other hand, RAN Engine [21] proposes a microservices-based approach to RAN customization.

While the literature cited above is varied in terms of its scope, it also shares a common goal in terms of seeking to perform RAN orchestration by leveraging the O-RAN RICs and 3GPP-centric base stations. However, RAN orchestration, while necessary, is not sufficient for network performance optimization in itself, being limited by the capabilities of the underlying base stations. For example, the framework presented in [20] requires explicit modifications to the 3GPP MAC layer to allow for RIC-based control over the resource allocation process. As noted in Section I, such limitations are often an outcome of the fact that cellular base stations are primarily self-contained entities not designed for external programmability, which is in stark contrast to the core tenets of O-RAN. To that end, there is a need for programmable base stations that are purpose-built for O-RAN, and HexRAN has been designed to address this goal.

## III. Key Requirements and Design Challenges

In this section, we identify the key requirements associated with designing a programmable base station that seamlessly blends the O-RAN and 3GPP ecosystems, followed by the design challenges that need to be overcome in order to achieve this goal.

### *A. The Tenets of Base Station Programmability*

The tenets of base station programmability are a set of four key requirements that serve as guiding principles in our pursuit of designing HexRAN. These include (i) dynamic RAN composability, (ii) programmable telemetry, control, and assurance, (iii) cross-functional adaptability, and (iv) an evolutionary approach to standards. The aforementioned tenets not only address technical systems-level considerations, but also highlight critical practical aspects that influence in-field RAN deployments.

To begin with, we note that while O-RAN has introduced the concept of disaggregation, the same level of disaggregation is not optimal for all end-user services and applications. For example, mobile broadband services with their high throughput requirements are best provisioned on central cloud NFs with sufficient capacity, while delay-sensitive metaverse services are ideally suited to edge NFs. However, networks today typically associate all users in a given cell with the same set of NFs, despite heterogeneous traffic requirements, thereby resulting in a sub-optimal one-size-fits-all experience. Instead, base stations should support the notion of **dynamic RAN composability**, wherein cells consist of a fluid set of service-optimized NFs that are cognizant of changing traffic demands.

Having identified a service-centric approach to RAN composition as a key tenet, we then consider the fact that base station programmability implicitly implies support for **programmable telemetry, control, and assurance**. Within this context, programmability not only refers to the ability to dynamically adjust statistics, configuration options, and key parameters but also extensibility in terms of being able to define new metrics and control procedures with ease. Interested readers may refer to our previous work [22] for additional context on programmability and associated requirements specific to the telemetry domain.

For decades, cellular standards have been primarily governed by a singular body, i.e., the 3GPP. However, with the advent of O-RAN, the O-RAN Alliance has come to play a key role in standardization too. The 6G and beyond era might witness the emergence of yet another standards organization. To that end, as varied standardization entities come to govern different aspects of the cellular architecture, it becomes imperative that base stations incorporate the notion of **cross-functional adaptability**, wherein the base station architecture is able to simultaneously cater to requirements set forth by the 3GPP, the O-RAN Alliance, and others. These requirements can often be orthogonal and even conflicting at times, further complicating the design. As noted earlier, the absence of such capabilities in the prior art is a major reason for sub-optimal NPO routines.

In a similar vein, as standards evolve to support an ever-increasing set of new features, the base station architecture should imbibe an **evolutionary approach to standards**. A typical case in point relates to O-RAN SMs, which are updated on a more frequent basis compared to the rest of the O-RAN and 3GPP standards. Within this context, an evolutionary approach to standards implies that the base station architecture should be sufficiently modular to allow for expeditious incorporation of new SMs, thereby resulting in a reduced time to market for the latest standardized capabilities, from a systems design and implementation perspective.

### *B. Design Challenges*

The realization of a programmable base station that conforms to the ideals set forth in §III-A poses a number of challenges within the context of the existing state-of-the-art.

- **Coarse-grained Network Function Disaggregation:** Thus far, the cellular base station has either consisted of a single NF, i.e., the gNB, or a collection of disaggregated NFs, i.e., RU, DU, CU-CP, and CU-UP. However, this coarse grained approach to disaggregation is a major impediment in realizing cell traffic-driven dynamic RAN composability.
- **Rigid Nature of the Cellular Protocol Stack:** The cellular protocol stack presents a well-defined control plane in the form of the Radio Resource Control (RRC) protocol, governed by the 3GPP. On the other hand, O-RAN espouses the use of external controllers for driving RAN operations, thereby resulting in a mismatch between

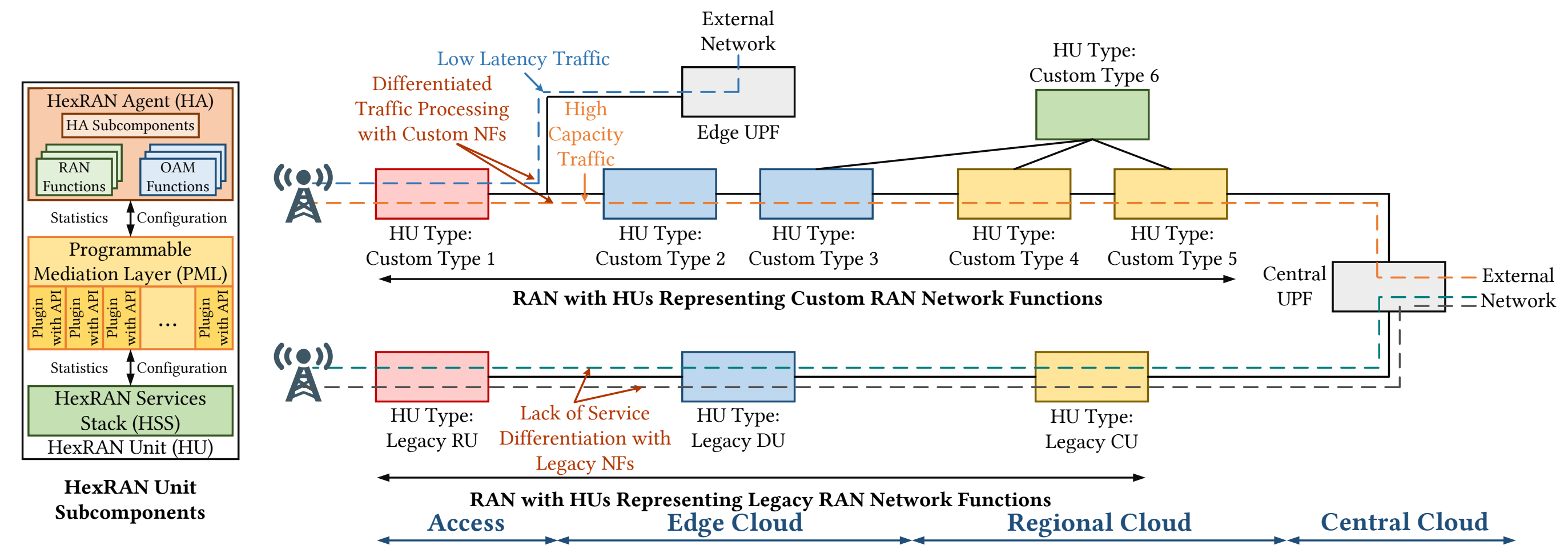

(a) HexRAN unit architecture.

(b) HexRAN-based RAN deployment showcasing both legacy and custom NFs.

Fig. 3: The HexRAN architecture.

TABLE I: Service-based decomposition of the RAN protocol stack.

| RAN Protocol Layer | HSS Services |
|---|---|
| **PHY** | High-PHY, Low-PHY |
| **MAC** | User Scheduling, Resource Allocation,and Prioritization, Channel Mapping, Multiplexing and Demultiplexing, Error Correction |
| **RLC** | Buffering and Retransmission, Segmentation and Reassembly, Duplicate Detection, Status Polling and Reporting |
| **PDCP** | Sequence Number Maintenance, Header Compression and Decompression, Integrity Protection and Verification, Ciphering and Deciphering, Reordering, Duplicate Discarding |
| **SDAP** | QoS Mapping and Steering, Reflective QoS Processing |
| **RRC** | Cell, User, and Bearer Management and Control |

O-RAN's goals and a RAN protocol stack that is not built for external control

- **Perceived Complexity of the O-RAN Specification:** O-RAN introduces several new interfaces, each characterized by a unique set of specifications, messages, and actions, ultimately leading to the perception that O-RAN adds significant complexity to an already complex RAN [23]. Therefore there is a need for solutions that simplify the integration of new features and enhancements within the base station.

## IV. HexRAN System Architecture Design

The overall HexRAN system architecture has been showcased in Fig. 3. As shown in the figure, RAN services are provided by a collection of NFs known as HexRAN Units (HUs). Illustrated in Fig. 3a, the HU is the primary building block of the HexRAN architecture, with each HU consisting of an underlying HexRAN Services Stack (HSS), along with a Programmable Mediation Layer (PML) and a HexRAN Agent (HA). Not all HUs incorporate the same set of HSS services, resulting in different HU types as shown in the figure. In the following sections, we detail each of these components along with their implications for the access network.

### A. *The HexRAN Services Stack*

Designed to overcome challenges relating to coarse-grained NF disaggregation, the HU eschews the traditional layered 3GPP protocol stack, and instead relies on a service-based approach through the HexRAN Services Stack (HSS). We build the HSS by decomposing 3GPP protocols such as the PHY, MAC, RLC, PDCP, RRC, and SDAP into a set of services that represent functionality that is expected from the aforementioned protocols as shown in Table I. For the sake of brevity, the table provides an illustrative example of key HSS services and should not be considered exhaustive. While the table provides a fine-grained decomposition of the MAC and other higher-layer protocols, we adopt a coarser approach to PHY decomposition by leveraging the existing O-RAN 7-2x fronthaul split to partition the PHY layer into high-PHY services and low-PHY services. The reason for this follows from the fact that intra-PHY communications pose strict throughput and latency requirements on the underlying interface [24], thereby precluding highly disaggregated PHY deployments.

The HSS provides unprecedented flexibility in terms of RAN composability. A given HU can contain one or more services from the HSS, and one or more HUs can work in tandem to form a complete base station. Based on the services they include, HUs can represent legacy 3GPP RAN NFs such as gNBs, CU-CPs, CU-UPs, DUs, and RUs, while at the same time, providing the opportunity to define custom RAN NFs. For example, by creating an HU which only includes "QoS Mapping and Steering" services, we can define a new RAN NF which functions as a gateway for the RAN, responsible for steering incoming user plane traffic from the core to the next hop RAN NF in the user plane processing chain.

Ultimately, the HSS serves as an enabler for the key tenet of dynamic RAN composability. As shown in Fig. 3b, a typical HexRAN-based RAN, consists of HUs representing both legacy as well as custom RAN NFs. The decoupled nature of HSS services allows for differentiated user traffic processing within the RAN, in line with application-specific requirements. The RAN provisioned in the figure consists of six HUs– HU Type 1 is deployed on compute-constrained access infrastructure, with the remaining HUs being deployed on a mix of edge and regional cloud infrastructure with successively higher compute capacity. As shown in the figure, traffic with strict latency requirements is processed by HU Type 1 which contains the complete set of HSS services, while traffic with high capacity requirements only leverages PHY services provided by HU Type 1.

The high capacity traffic is then further processed by HU Types 2 through 5 with sufficient compute availability for the processing of throughput-heavy traffic. HU Type 6 provides control plane services and is thus not involved in the processing of user traffic. In this manner, the compute-constrained HU 1 can preserve its resources for low-latency traffic, while providing minimum feasible compute, i.e., PHY-only services, to other traffic types. The distribution of traffic across HUs can be done dynamically on a per-user, per-slice, or per-bearer basis, to allow for maximum operational flexibility. Fig. 3b also showcases a legacy RAN provisioned using HUs, however, in this case, the legacy RAN NFs offer significantly lower flexibility, with all traffic being processed in an identical manner across the RU, DU, and CU.

On the other hand, an HU-driven RAN can scale in an extremely efficient manner by adding or removing additional compute for or instances of only those HUs which contain services that need additional capacity. Such scaling is typically driven by variations in cell traffic demand. For example, in a legacy RAN, providing additional compute capacity to the scheduler entails scaling of the entire DU, however, with HexRAN, the CSP need only scale the user scheduling service, thereby providing a more efficient utilization of compute resources. Similarly, HexRAN's ability to dynamically split traffic across HUs with varying levels of HSS services also allows for targeted load balancing across the RAN in accordance with end-user traffic variations.

### B. The Programmable Mediation Layer

Having introduced the HSS, we now detail the next key component within the HU– the Programmable Mediation Layer (PML). The PML has been designed to address key requirements relating to programmable telemetry, control, assurance, and cross-functional adaptability while attempting to overcome challenges arising from the rigid nature of the cellular protocol stack. The PML follows a plugin-based architecture, with each plugin providing a set of specific APIs that can interact with the underlying HSS to configure RAN operations and collect system statistics. These APIs are made available to the HexRAN Agent for enabling O-RAN-driven operations, as illustrated in Fig. 3a. To that end, the PML can be considered as an adaptation layer that bridges the 3GPP and O-RAN domains by providing a functional abstraction of the RAN protocol stack.

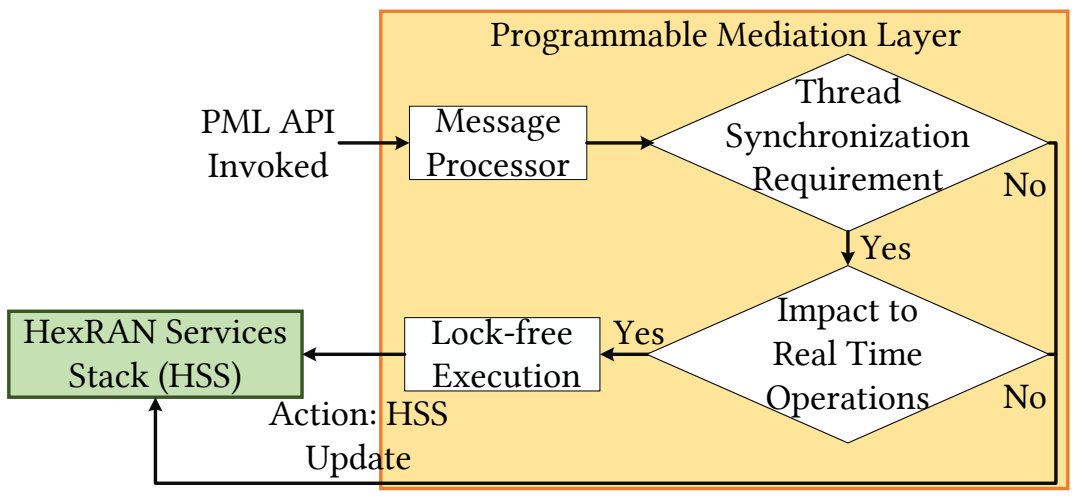

Fig. 4: Lock-free PML API execution framework.

While the primary goal of the PML is to simplify O-RAN-related operations within 3GPP RANs, HexRAN also lays a strong emphasis on API programmability and reusability. For example, if we consider a PML-provided Link Adaptation Framework, the corresponding APIs can be leveraged by an xApp providing energy optimization features (through a corresponding O-RAN RAN function within the HexRAN Agent), an rApp providing throughput optimization (through a corresponding YANG model and OAM function within the HexRAN Agent), or even a closed loop power control service provided by the HSS. Furthermore, we note that the abstractions provided by the PML are key enablers for a third-party xApp/rApp ecosystem. For example, the vendor designing the base station can provide PML APIs that enable programmable control of the MAC scheduler, which can then be leveraged by other xApp/rApp developers to deliver optimized user scheduling, without either entity having to expose their proprietary business logic.

In the interest of conflict-free operations, HexRAN does not allow external entities to interact with the HSS directly by design. All HSS-related interactions must go through the PML, which includes a built-in conflict mitigation mechanism. In the current implementation of HexRAN, each PML API uses a FIFO scheduler that queues calls for execution in order of their respective arrivals. Furthermore, there also exists a lockout period which prevents the same parameter from being modified in quick succession by different entities leveraging the same PML API.

Since the PML interfaces with the HSS, it is critical that API calls do not disrupt real-time RAN operations such as scheduling. However, the multi-threaded nature of HexRAN necessitates the need for thread synchronization to execute configuration changes. To that end, as shown in Fig. 4, the PML APIs within HexRAN incorporate a lock-free synchronization mechanism to ensure minimal disruption to real-time operations. As shown in the figure, when a given PML API is invoked by an O-RAN RAN function or any other entity, if the system detects a need for thread synchronization that could potentially impact operations, it executes the resulting action in a lock-free manner.

In summary, through the PML, HexRAN effectively eliminates the need for other base station functions (e.g., O-RAN RAN functions, OAM functions, etc.) to directly interact with the HSS, thereby helping overcome the inflexible nature of cellular protocols. Additionally, the PML provides several

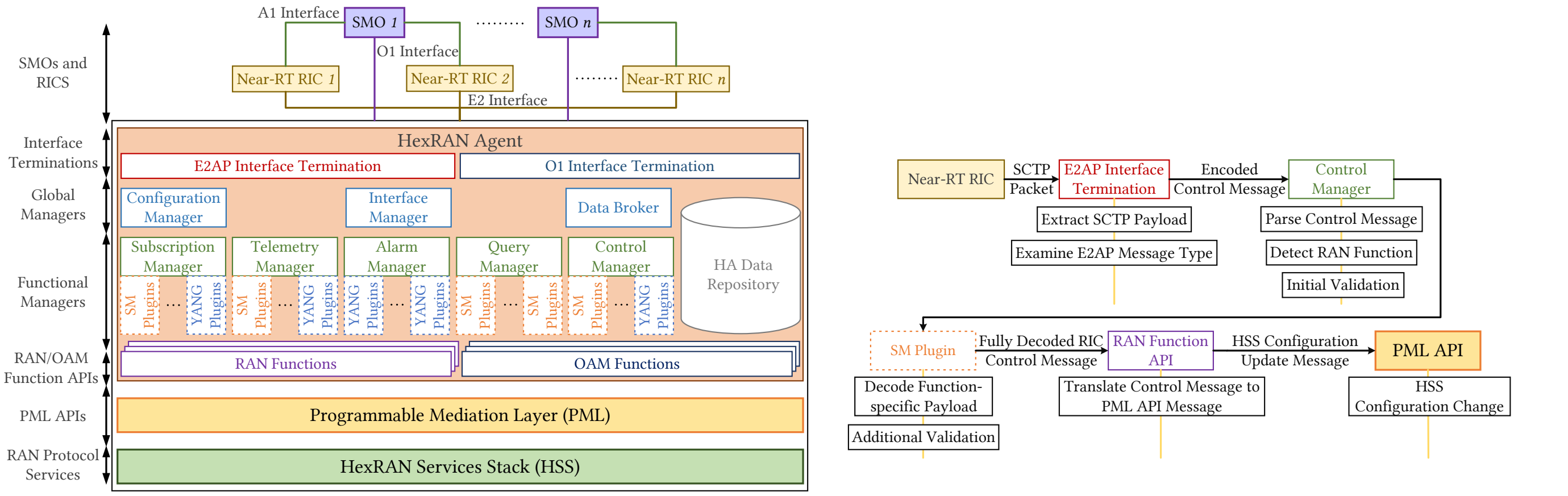

(a) HexRAN Agent architecture and integration within the HU. (b) RIC control message processing within HexRAN Agent.

Fig. 5: HexRAN Agent architecture and message flow.

advantages including: (i) simplified design and seamless integration of new base station functions, (ii) native support for multi-vendor RAN-RIC solutions, and (iii) conflict-free operations.

### C. The HexRAN Agent

The HexRAN Agent (HA) is responsible for all RAN-level interactions of an HU with other external non-HU NFs. An integral part of the HU, key responsibilities of the HA include setup and maintenance of the O1 and E2 interfaces, execution of statistics and control requests from the SMO and RIC, and delivery of telemetry and alarms back to the RICs. At the northbound, the HA interfaces with the SMO and near-RT RIC over the O1 and E2 interfaces respectively, while at the southbound, the HA interacts with the PML. The HA represents a first-of-its-kind nerve-center for O-RAN-related operations within the base station, and, consequently, we incorporate the following principles in the design of this critical component.

- **Architectural Modularity:** Since O-RAN is a fast evolving field, the HA should incorporate a modular design with replaceable subcomponents.
- **Feature Extensibility:** A large part of the HA's operations are centered around RAN and OAM functions. Therefore, the HA architecture must be extensible by design to allow for seamless integration of new functions and their accompanying service or data models.
- **High-performance Message Processing:** In dealing with near-RT operations, both message processing latency and reliability are paramount, therefore, the HA should minimize processing latency while maximizing reliability.
- **Multi-SMO/RIC Support:** The HA should be able to interface with multiple SMOs and near-RT RICs. Support for multiple RICs not only enhances operational resiliency but also enables advanced use cases such as RAN sharing, which allow for differentiated control of a common RAN infrastructure.

The HexRAN Agent architecture along with its implementation within the HexRAN Unit has been shown in Fig. 5a. As illustrated in the figure, the HA includes several classes of subcomponents– interface terminations, global managers, functional managers, and an HA data repository, along with a set of RAN functions and OAM functions. More specifically, the HA incorporates two interface terminations– an E2 termination and an O1 termination. The E2 termination incorporates an SCTP stack and provides support for the encoding and decoding of E2AP messages, while the O1 termination provides a NETCONF server which is used by the SMO's NETCONF client to interact with HUs.

Next, the global managers represent those subcomponents which are responsible for general administrative and management tasks. The HA includes three such managers– the Configuration Manager, the Interface Manager, and the Data Broker. The Configuration Manager is responsible for parsing O1 and E2-related aspects of the RAN configuration, e.g., information about the RAN and OAM functions supported by a given HU, list of authorized NETCONF clients (e.g., SMOs), as well as reachability information concerning near-RT RICs. We note that in keeping with O-RAN specifications, RAN functions are used for RAN-near-RT RIC interactions, while OAM functions are used for RAN-SMO interactions. Since HexRAN requires that RAN and OAM functions interact with the HSS through the PML only, there exists a dependency between these functions and the PML's plugins. To that end, the Configuration Manager also performs dependency verification and considers only those functions for which the required PML plugins are present in the RAN. The Configuration Manager then leverages the Data Broker to update the HA Repository with the parsed function data and SMO/RIC-related context.

The Interface Manager handles all interface management procedures for the E2 and O1 interfaces. Such actions are performed based on the RIC and SMO data present in the HA Repository. The repository maintains a separate context for each RIC and SMO, and based on a given RIC or SMO's response, the Interface Manager activates the requested RAN (or OAM) functions for interactions with that RIC (or SMO) instance. More generally, once a specific set of resources within the RAN are associated to a given RIC (or SMO) instance, other instances cannot exercise the control over those resources to avoid conflicts. The Data Broker provides read-write access to the HA Repository. As mentioned above, in addition to maintaining RIC and SMO-related context, the repository also includes information regarding the RAN state in terms of configuration and operational data. The RAN state-specific content of the repository is governed by the Service Models (SMs) and YANG models active within the RAN.

On the other hand, the functional managers perform RAN and OAM function-specific operations. The HA provides five functional managers– the Subscription Manager, the Telemetry Manager, the Alarm Manager, the Query Manager, and the Control Manager. Key responsibilities include the ability to parse RAN and OAM function-specific message content and invoke the corresponding RAN/OAM function APIs. The Subscription Manager is responsible for handling telemetry requests from the RIC and SMO, while the Telemetry Manager is responsible for collecting and delivering statistics. Similarly, the Alarm Manager raises alarm notifications, while the Query Manager is designed to handle the `RIC Query` procedure. Finally, the Control Manager is responsible for parsing configuration updates and RAN parameter control requests from the RIC (and SMO), and invoking the relevant RAN and OAM function APIs for the execution of said requests. We note that in accordance with the E2AP and O1 specifications, procedures involving the near-RT RIC leverage the Subscription, Telemetry, Control, and Query Managers, while procedures involving the SMO leverage the Subscription, Telemetry, Control, and Alarm Managers.

For example, for the standardized RAN Control (RC) SM, the various functional managers within the HA are responsible for the execution of RC-specific control and policy directives from the RIC and delivery of RC-specific telemetry and query data back to the RIC. However, since it is not practically feasible to deploy dedicated functional managers per RAN (or OAM) function, HexRAN introduces the concept of E2SM and YANG plugins as shown in the figure. Each RAN (or OAM) function within HexRAN includes a set of SM (or YANG) plugins for the functional managers, which are used to decode function-specific payloads. More specifically, the Subscription, Telemetry, and Control Managers support both SM and YANG plugins, the Query Manager supports only SM plugins, and the Alarm Manager supports only YANG plugins. This distinction follows from the fact that interactions with the near-RT RIC are governed by Service Models, while interactions with the SMO are governed by YANG models. A plugin-based approach imparts substantial extensibility to the HexRAN Agent, while also simplifying RAN and OAM function design. With HexRAN, function designers need only provide the supporting plugins and RAN/OAM function APIs without having to deal with the complexity associated with messaging over the E2 and O1 interfaces.

In order to further describe the functioning of the HA, we consider the following illustrative example within the domain of RAN-RIC interactions. Upon receiving an SCTP message from a given near-RT RIC, the HA's E2 termination extracts the SCTP payload, i.e., the E2AP message, examines the message type, and offloads it to one of the functional managers. The offload mechanism is implemented through a set of function callbacks which invoke the appropriate manager for further processing. For example, as shown in Fig. 5b, if the incoming request belongs to the `RIC Control` procedure, the E2 Interface Termination will offload it to the Control Manager. The Control Manager then parses the E2AP payload and performs an initial validation step. Upon successful validation, the RAN function-specific payload within the E2AP message is further offloaded to the appropriate SM plugin. The SM plugin decodes the RAN function-specific payload, i.e., the `RIC Control Header` and `RIC Control Message`, and may perform additional validation steps. Finally, this decoded payload is used to invoke the corresponding RAN function's API which translates the decoded message to an appropriate PML API message. The RAN function API then invokes the appropriate PML API which finally executes the requested control action.

Similarly, we also consider the following RAN-SMO interaction. Upon receiving an `Edit Config` message from the SMO, the O1 termination within the HA invokes the Data Broker, which performs an initial validation of the data format and commits the configuration to the HA Data Repository, upon which the Control Manager receives a notification regarding a change in the RAN configuration data. The YANG plugin of the Control Manager then extracts and validates the provided configuration, and invokes the corresponding OAM function API to execute the configuration (through the PML).

The HA thus ensures that HexRAN is able to seamlessly adapt to varying degrees of standards evolution, while overcoming challenges arising from the complexity of specifications. More generally, each of the three components that constitute an HU play a critical role in achieving the tenets of base station programmability identified in §III-A while addressing the challenges delineated in §III-B.

## V. Vertically Integrated RAN Programmability

With the versatile HU as a building block, HexRAN serves as the perfect fit for a variety of use cases that require RAN programmability. Within the context of this paper, we focus on RAN slicing. With its ability to support multiple logical isolated services over a common physical infrastructure, RAN slicing has been recognized as a flagship use case for the O-RAN architecture [25]. Beyond O-RAN, RAN slicing has also witnessed significant traction within the research community with an emphasis on solutions that enable the optimized allocation of resources across slices [19], [26]. As networks evolve towards 6G, slicing is expected to support a plethora of new and emerging applications such as extended reality and the metaverse, cloud gaming, and vehicular communications, with each presenting a diverse set of requirements, requiring robust network programmability, along with closed-loop automation and control. To that end, in this section, we showcase how HexRAN can be used in conjunction with the near-RT RIC to deliver vertically integrated RAN programmability in support of network slicing.

Referring back to the HexRAN architecture in §IV, we note that the HSS is agnostic to network slicing, with the choice of HSS services in a given HU being solely governed by network deployment, compute availability, and transport capacity considerations. However, both the PML and HA have a key role to play in enabling RAN slice programmability. More specifically, through this section, we introduce the FlexiSlice Framework (FSF) which provides programmable RAN slicing-specific plugins for the PML and HA components. A complete deployment of FlexiSlice within the HexRAN Unit has been showcased in Fig. 6. From the figure, we note that the FSF PML plugin provides the FlexiSlice Context, the FlexiSlice Scheduling Framework, and supporting APIs, while the FSF-specific HA components include the FlexiSlice O-RAN service model with supporting plugins and FlexiSlice RAN function as detailed in the following sections.

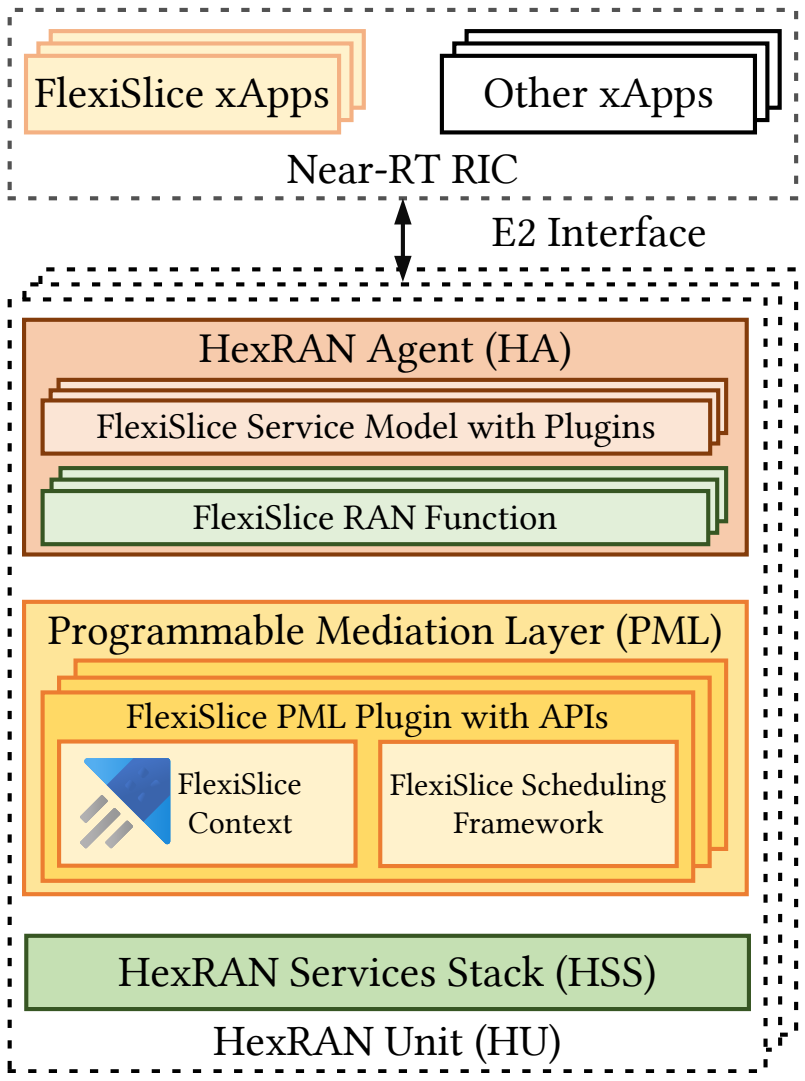

Fig. 6: The FlexiSlice framework as deployed within the HexRAN Unit.

### *A. FlexiSlice Context*

The FlexiSlice (FS) Context serves as the data repository for RAN slicing-related operations within an HU. It stores information pertaining to both RAN slice configuration and statistics. As a part of the PML, external entities can interact with the FS Context to obtain telemetry data and configure RAN slices through a set of FS PML APIs. In the interest of maximizing flexibility and programmability, FS introduces a novel RAN slice definition based on the following parameters.

- **Operational State:** A RAN slice within FS can be characterized by its operational state. At any given point in time, a slice can exist in one of five states– idle, dedicated, prioritized, shared, and hybrid, as described below.
  - Idle State: A RAN slice in the idle state does not have any data sessions (or more specifically data radio bearers, i.e., DRBs) associated with it. An idle slice does not consume any RAN-level radio resources.
  - Prioritized State: A slice in the prioritized state has priority access to the resources assigned to it, i.e., should the slice need those resources, no other slice can preempt it. However, resources that are not used by the slice ultimately enter the shared pool, from where they can be used by other slices.
  - Shared State: A slice in the shared state does not have an explicit resource assignment. Instead, all slices in the shared state have access to a shared resource pool. Resource allocation from the shared pool to shared slices is done in accordance with the specific shared slice's priority.
  - Hybrid State: A slice in the hybrid state incorporates all three characteristics described above, i.e., a hybrid slice can be viewed as a logical construct that includes all three states (dedicated, prioritized, and shared) in the form of sub-slices.
- **Radio Resource Configuration:** Each slice is further characterized in terms of its radio resource configuration. More specifically, the resource configuration parameters depend on the state of the slice.
  - Dedicated Slices: For slices in the dedicated state, the radio resource parameter defines the number of resource blocks that are dedicated to that slice.
  - Prioritized Slices: For slices in the prioritized state, the radio resource parameter defines the number of resources blocks that the slice has priority access to. The prioritized resource assignment can be viewed as an upper bound on the radio resource availability for that slice.
  - Shared Slices: Since for slices in the shared state, there is no explicit radio resource assignment, the resource parameter represents a priority value which defines the level of priority the given slice has in terms of access to the shared resource pool.
  - Hybrid Slices: Hybrid slices are characterized by the presence of a dedicated radio resource configuration, a prioritized radio resource configuration, and a shared priority parameter.
- **Frequency-domain Scheduling Algorithm:** Driven by a use case-specific perspective, FS is designed to allow custom slice-specific scheduling algorithms, therefore, the frequency-domain scheduling algorithm is a key slice characterization parameter.
- **HU Associations:** In the interest of ensuring optimal compute utilization and delivering the required throughput and latency, FS RAN slices are also characterized by the HUs associated with the slice. An approach of this kind ensures that the best candidate HUs are always selected for the specific services supported by that slice.

Working in tandem, the parameters defined above ensure that FS RAN slices can be customized to meet the targeted requirements of different use cases, a key feature that is notably absent in the prior art. Furthermore, the varied slice states ensure that FS also achieves full compliance with 3GPP's Network Resource Model [27]. Within that context, in order to ensure seamless transitions between difference slice states, FS introduces a novel RAN slice state machine as illustrated in Fig. 7.

From Fig. 7, we note that when a RAN slice is initialized within HexRAN, it is in the idle state with no active DRBs associated with it. The slice is configured with a default state to which it can transition upon DRB addition for any data sessions associated with that slice. Once a slice transitions to either one of the dedicated, prioritized, shared, or hybrid states, it remains in that state unless either an explicit state change request is issued through the FS PML API or if the number of DRBs associated with that slice reaches zero. In the event that the latter condition is met, slices in the prioritized and shared states switch to the idle state with the default active state being set to shared. On the other hand, hybrid slices switch to the dedicated state while retaining their dedicated resource assignment only, the remaining resources associated with this slice now enter the shared pool. Finally, dedicated slices experience no state change in the event that the number of DRBs reach zero.

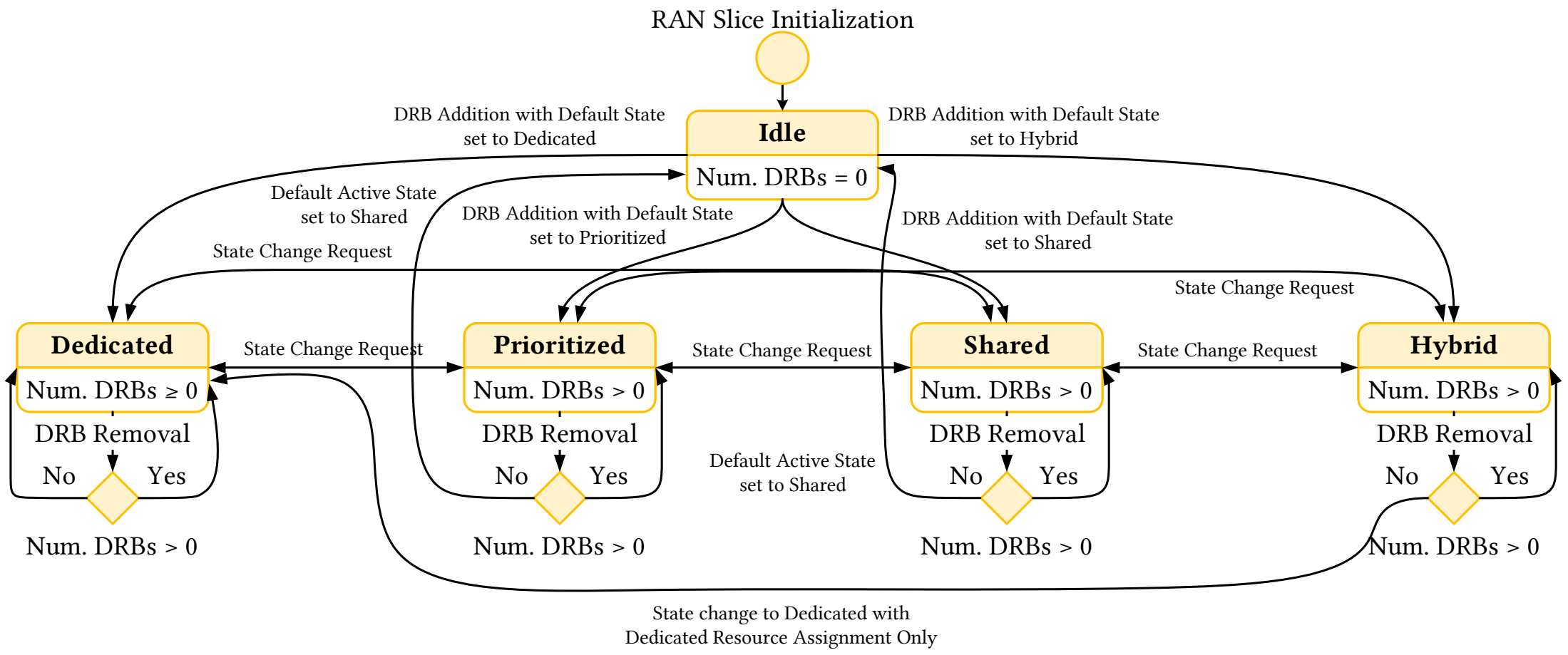

Fig. 7: The FlexiSlice state machine.

In this manner, the FS state machine ensures efficient management of radio resources to reflect the operational conditions being experienced by the network. For example, during regular operational conditions, public safety slices could operate in the prioritized state, but during emergencies, such slices could be switched to the dedicated state in order to ensure that even as public safety devices join and leave the network, such devices always have a guaranteed set of resources available to them.

The operation of the aforementioned state machine as well interactions with the FS Context are governed by the FS PML APIs. In line with the HexRAN design philosophy, these APIs are intended to be used by O-RAN RAN and OAM functions to execute control directives as well as obtain telemetry data. More specifically, FS provides the following PML APIs.

- **FS Telemetry Registration Request:** Used to register statistics requests (sent by xApps on the near-RT RIC) with the FS Context. As part of the request, the entity invoking the API needs to provide the target resources, i.e., slices and/or users (i.e., UEs), along with a corresponding telemetry trigger, which can be either periodic or event-specific.
- **FS Statistics Request:** Upon the detection of an event trigger, the FS Statistics Request routine can be used to obtain statistics-related parameters associated with specific slices and/or users.

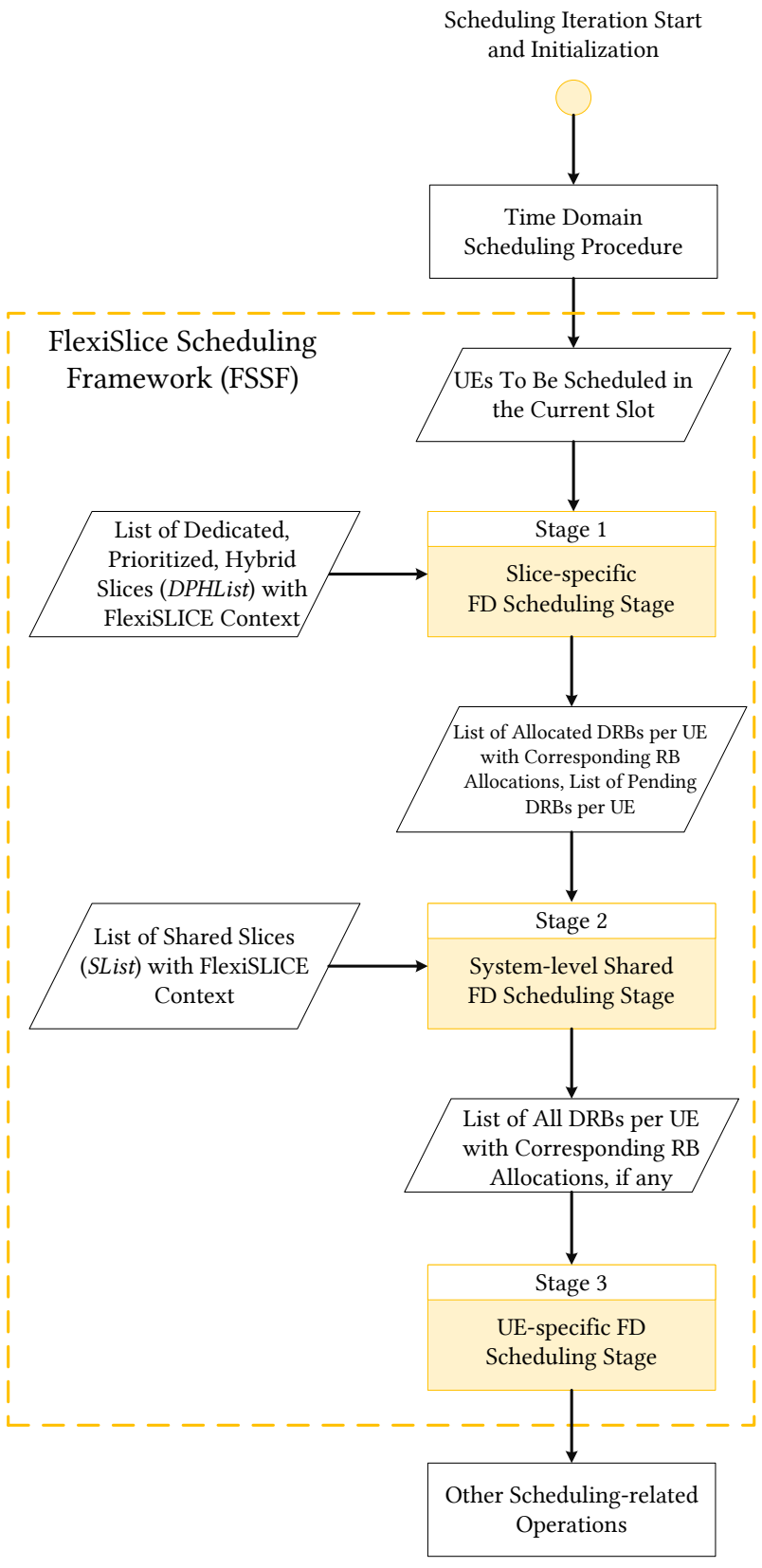

Fig. 8: The FlexiSlice Scheduling Framework.

- **FS Context Change Request:** Upon detection of an event trigger, the FS Context Change Request is used to obtain specific information about the 3GPP or non-3GPP procedure responsible for changes to the FS Context. This procedure ensures that external entities such as the RIC are notified about slice-related changes that occur within the RAN.
- **FS Control Request:** Used to adjust parameters within the FS Context relating to one or more slices and/or users. The parameters are based on the aforementioned slice characterization.

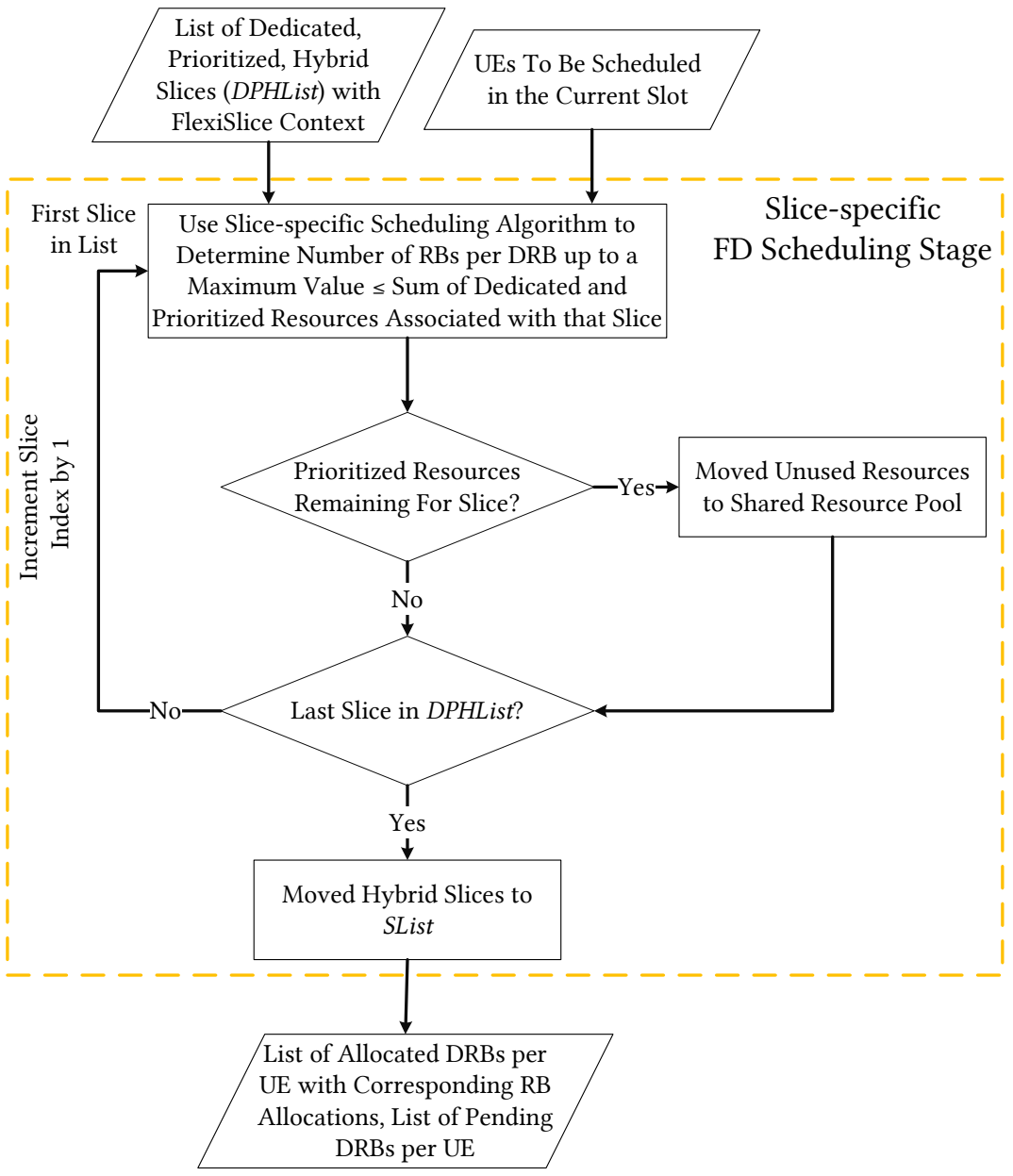

Fig. 9: Slice-specific frequency domain scheduling stage.

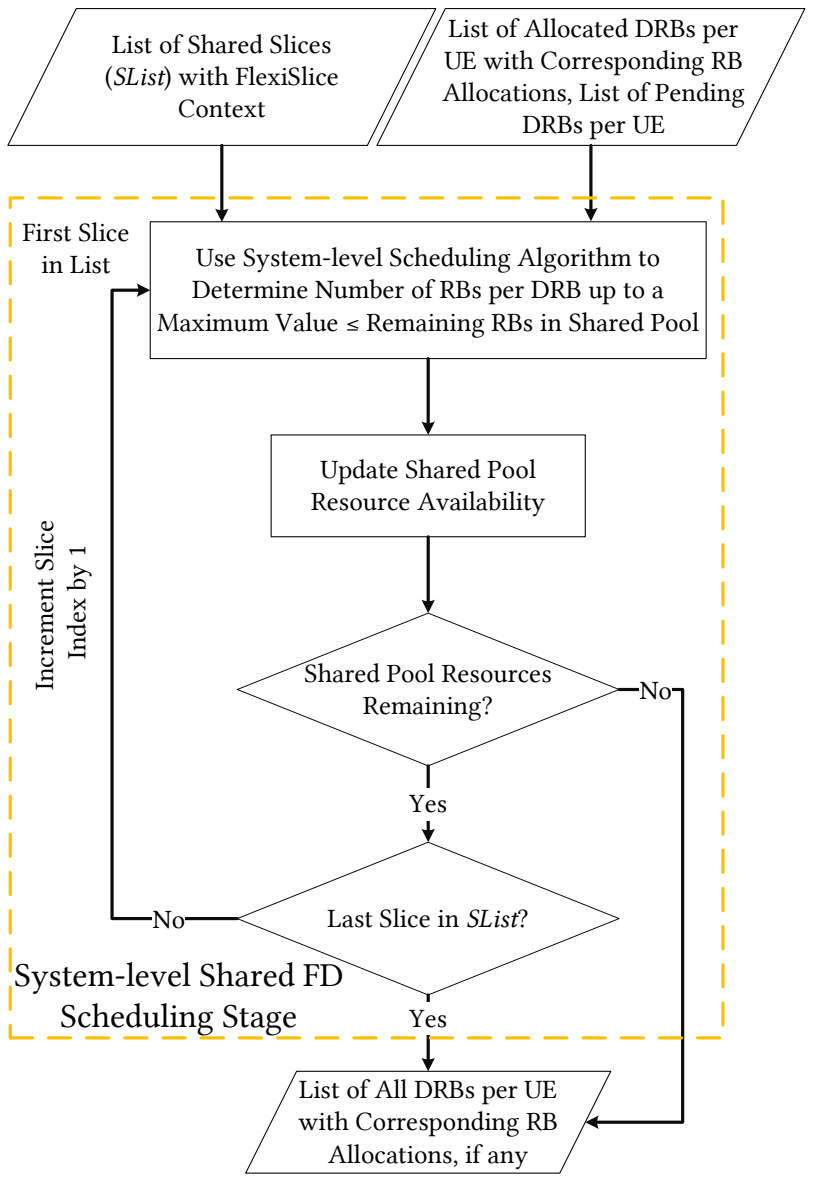

Fig. 10: System-level shared frequency domain scheduling Stage.

### B. FlexiSlice Scheduling Framework

The FlexiSlice Scheduling Framework (FSSF) is another key aspect of the FS PML plugin. The FSSF is responsible for allocating frequency domain (FD) radio resources to UEs in a slice-aware manner on a per-transmission time interval (TTI) basis. As shown in Fig. 8, the FSSF operations begin after the time domain scheduling process has been completed and the list of UEs to be scheduled for that TTI is available. The FSSF incorporates a three-stage FD scheduling process as showcased in the figure. By the end of the third stage, all eligible UEs have been allocated virtual resource blocks (VRBs), followed by existing scheduler-specific operations. Interested readers may refer to [28] for additional information regarding VRBs. The three stages are detailed below.

*1) Slice-specific Frequency Domain Scheduling Stage:* As detailed in Fig. 9, the slice-specific FD scheduling stage is used for allocating radio resources to the dedicated, prioritized, and hybrid slice types. This stage takes as input the list of UEs to be scheduled in the current slot along with the list of dedicated, prioritized, and hybrid slices (*DPHList*) characterized by slice-specific context information. The *DPHList* can be sorted in any suitable manner. Then, based on the slice-specific scheduling algorithm, the FSSF Stage 1 process determines the number of RBs to allocate to each DRB associated with a given slice. In general, the sum total of RBs allocated across DRBs within a given slice cannot exceed the sum of dedicated and prioritized resources associated with that slice. After the resource amount has been determined, any remaining prioritized resources are moved to the shared pool, and the process repeats until all slices in *DPHList* have been exhausted, post which, the hybrid slices are moved to the list of shared slices, i.e., *SList*, and Stage 2 begins. In this manner, *SList* contains both shared slices as well as hybrid slices for which the dedicated and prioritized resource assignment has already been scheduled.

*2) System-level Shared Frequency Domain Scheduling Stage:* The system-level shared FD scheduling stage has been showcased in Fig. 10. As shown in the figure, the inputs to this stage include the list of shared slices (*SList*), along with the relevant slice-specific context, list of allocated DRBs per UE with corresponding RB allocations, and a list of pending DRBs per UE. Slices within *SList* are sorted in accordance with their shared priority parameter. The FSSF Stage 2 leverages a common system-level scheduler to determine the amount of radio resources to allocate to DRBs associated with each *SList* slice up to a maximum value bounded by the resource availability within the shared resource pool. Stage 2 ends when either all slices within *SList* have been allocated or the shared pool is exhausted.

*3) UE-specific Frequency Domain Scheduling Stage:* While FSSF Stages 1 and 2 are primarily intended for determining the absolute number of RBs to be allocated to DRBs across slices, Stage 3, i.e., the UE-specific FD scheduling stage, is responsible for assigning specific VRBs to UEs in accordance with the resource allocation decisions taken during Stages 1 and 2. For example, if the sum total allocation for a given UE, as determined by Stages 1 and 2, is 20, then Stage 3 determines that VRB5 through VRB24 should be allocated to that UE.

In this manner, the FSSF ensures use case-specific customization for RAN slicing, with each RAN slice being able to use its own custom scheduling algorithm for allocating frequency domain radio resources to its constituent bearers. For example, a given slice could use proportional fair scheduling, a second slice could use round robin scheduling, while a third slice could use maximum throughput scheduling. We note that the introduction of new slice-specific and system-level scheduling algorithms is not the goal of the FSSF, instead, our focus is on providing a framework that allows for flexible and programmable use of a variety of different scheduling algorithms [26], [29].

### C. FlexiSlice O-RAN Service Model and RAN Function

Having detailed the FS PML plugin, we now describe the FSF HA plugin which incorporates the FS O-RAN service model (SM) and RAN function. The FS service model governs the various interactions between the near-RT RIC and the FS RAN function, while the FS RAN function leverages the FS PML APIs in order to interact with the FS context, to achieve programmable slice control and assurance. In particular, the FS service model provides support for O-RAN's Report and Control services. Space limitations preclude the inclusion of additional background on the aforementioned services, however, interested readers may refer to the O-RAN specifications [30] for further information. In particular, the FS SM introduces the following Report services.

- **FS Slice Context Report:** The FS slice context report contains configuration information and statistics data for one or more RAN slices, including parameters such as the slice ID, state, HU associations, frequency domain scheduler, and radio resource configuration, along with a list of associated bearers. For each bearer, the report includes parameters such as bearer priority, throughput, packet delay, packet loss rate, and buffer occupancy.
- **FS UE Context Report:** The FS UE context report provides RAN slice-related configuration information and statistics data for one or more UEs, incorporating parameters such as the UE ID, modulation and coding scheme (MCS), channel quality indicator (CQI), and block error rate (BLER), in addition to a list of associated bearers containing the aforementioned bearer-specific parameters.
- **FS Context Change Report:** The FS context change report is required for synchronizing RAN slice-related context between the near-RT RIC and an HU. For slices included within the context change report, typical parameters include the triggering event responsible for the context change, along with a list of resulting outcomes describing changes to the slice-HU associations (e.g., addition or removal of HUs associated with the slice), scheduling algorithm (e.g., change in the frequency domain scheduling algorithm used by the slice), resource configuration (e.g., updates to the radio resource assignment for the slice), and associated bearer list (e.g., addition, removal, or modification of bearers associated with the slice).

In a similar vein, the FS SM also introduces the following Control services.

- **FS Slice Configuration Routine:** The FS slice configuration routine provides control of several RAN slice parameters such as the slice's state, the HUs associated with it, the specific frequency domain scheduler in use, and the associated radio resource configuration.
- **FS UE Configuration Routine:** The FS UE configuration routine allows for UE-relevant bearer-specific parameter updates such as those relating to bearer priority, thereby allowing for dynamic prioritization of specific bearers within the slice. Such control is a key enabler for operational service prioritization as users within the same slice are typically offered the same QoS.

In this manner, through the FS HA plugin, xApps on the near-RT RIC can obtain fine-grained RAN telemetry while also delivering programmable slice customization, which is enabled through a combination of the FS PML and HA plugins.

## VI. System Implementation and Evaluation

In this section we first present details concerning a systems-level implementation of HexRAN as well as its deployment on an over-the-air experimental testbed with COTS UEs. Then, by leveraging the tested, we conduct a comprehensive evaluation of HexRAN along three major themes– system scalability, agent performance, and RAN slice programmability.

### A. HexRAN Implementation

In order to implement a systems-level prototype of HexRAN, we first needed to select a 3GPP-compliant RAN protocol stack. We note that two open-source projects– OpenAirInterface (OAI) [31] and srsRAN [32] meet this requirement. To that end, we leveraged OAI as the base for realizing HexRAN. In order to build the HexRAN Services Stack, we decomposed the RAN protocol layers into a set of services as outlined in Table I. In general, interfaces between user plane HSS services were realized using the GPRS Tunneling Protocol (GTP), with the exception of the high-PHY northbound which leveraged the nFAPI [33] interface. On the other hand, HSS control plane interfaces were realized using a custom application protocol called HexAP deployed over SCTP. We omit a detailed description of HexAP in the interest of brevity, however, akin to other cellular application protocols like the F1AP [34] and NGAP [35], HexAP also provides specific interface, user, and bearer management procedures for inter-HexRAN Unit interactions.

Once the HSS component of the HU had been implemented, we built the Programmable Mediation Layer framework along with the FlexiSlice plugin for programmable RAN slicing. The PML was implemented in C and relies on two external dependencies–Concurrency Kit [36] and `liblfds` [37]–for achieving lock-free operations. The entire framework along with the plugin accounted for approximately 6000 lines of code, i.e., 6 kLOC. Next, we implemented the HexRAN Agent along with its various subcomponents from scratch. Inter-component communication within the HA was achieved through a combination of OAI's Intertask Interface (ITTI) library and nanomsg-NG (NNG) [38]. Each stage of the message processing pipeline was decoupled from the one before it through a system of message queues, ensuring that the message processing within the HA experiences minimal delays in the interest of low-latency operations. Furthermore, the HA was designed to host multiple instances of each global and functional manager as well as their respective plugins, allowing for load balancing during heightened RAN-RIC interactions. Also written in C, the entire HA implementation accounted for over 15 kLOC.

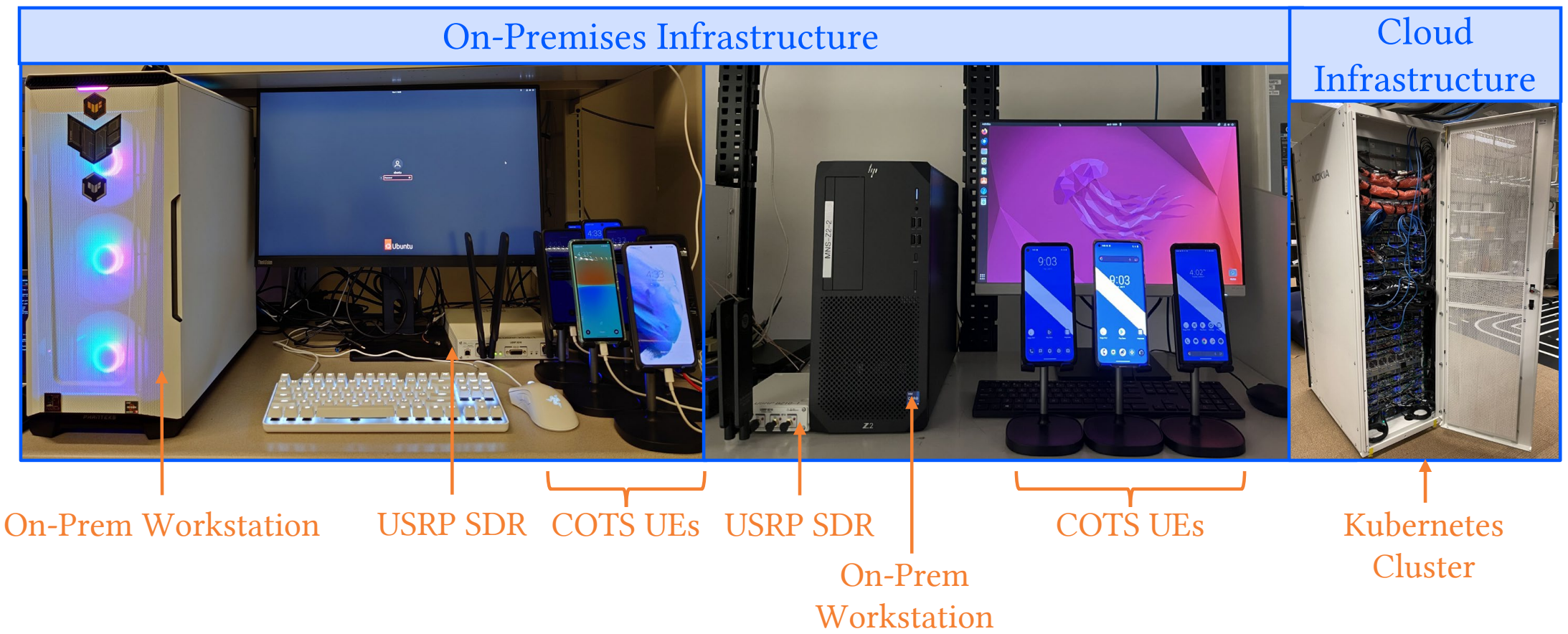

Fig. 11: HexRAN experimental testbed showcasing cloud and on-prem infrastructure.

Finally, in support of the programmable RAN slicing use case, we implemented two xApps– FS Telemetry and FS Configuration, on the HexRIC platform [39]. As their names suggest, the FS Telemetry xApp served as a key enabler for obtaining the RAN performance statistics showcased in the subsequent sections, while the FS Configuration xApp played a key role in executing different experimental scenarios used in the evaluation herein.

## B. *Experimental Testbed Setup*

In the interest of obtaining realistic and at-scale results we deploy HexRAN on an experimental testbed characterized by a mix of cloud and on-premises (on-prem) infrastructure as shown in Fig. 11. The cloud infrastructure consists of a Kubernetes cluster of eight nodes, while the on-prem infrastructure consists of two workstations, each attached to a USRP SDR. This infrastructure is used to create two test environments– a hardware radio environment and an emulated radio environment. In either case, both the core network (Open5GS [40]) and HexRIC platform are deployed on the cloud infrastructure. More specifically,

- **Hardware Radio Environment:** As shown in Fig. 11, the hardware radio environment is anchored by two USRP SDRs and multiple COTS UEs. Throughout all the over-the-air experiments, the HUs that contain the low-PHY and high-PHY HSS services are always deployed on the on-prem workstations, while all other HUs can be freely deployed across the on-prem and cloud infrastructure depending on the scenario under evaluation.
- **Emulated Radio Environment:** While the hardware radio environment is indispensable for conducting field-realistic experiments, large-scale evaluation scenarios often require several cells. In such cases, we turn to an emulated environment. At the outset, this large-scale test environment is *emulated* and not *simulated*, i.e., the entire HexRAN software stack (including the HSS, PML, and HA) is identical across both the hardware radio and emulated radio environments, with the only difference being that the emulated environment uses emulated radios (EMRs) and UEs provided by the *rfsim* component of OAI. The emulated environment is deployed on the cloud infrastructure and can host up to 50 cells and a total of 250 UEs. Furthermore, we note that the scale of the emulated environment renders manual experimentation infeasible, and therefore we leverage Ansible [41] to automate the experiments.

## C. *Enhancing RAN Scalability*

In comparing HexRAN to existing cellular base stations, we note that the HU architecture introduces several new components. While these components provide unprecedented flexibility, we recognize that they could also lead to an increase in the overall compute resource consumption. Therefore, it is vital to ascertain whether the HexRAN architecture introduces additional overhead both in terms of CPU and memory utilization, as well as its impact on the network's performance. To that end, we begin our experimental evaluation by contrasting HexRAN with the baseline 3GPP base station architecture, both in terms of compute utilization as well as throughput performance.

We conduct this experiment by deploying two base stations on the hardware radio environment– one representing HexRAN and the other representing the baseline, each broadcasting a single 40 MHz n78 carrier. The HexRAN base station consists of a collection of HSS services deployed on the same physical compute node, which are functionally equivalent to a 5G standalone (SA) gNodeB (gNB), while the baseline consists of a vanilla OAI gNB. Within this context, we compare the CPU and memory utilization as the number of users associated with each base station increases from 2 to 12, with each user engaged in a downlink traffic session of 10 Mbps. The results thus obtained have been showcased in Figs. 12a and 12b. From the figure, we note that both the HexRAN gNB and the baseline are largely identical in terms of their compute and memory requirements, showcasing that the HSS does not introduce additional compute overhead. Furthermore, as part of this experiment, we also compare the aggregate downlink throughput achieved by each gNB in Fig. 12c, noting that the HexRAN gNB matches the throughput achieved by the baseline, ranging from 20 Mbps for 2 users to

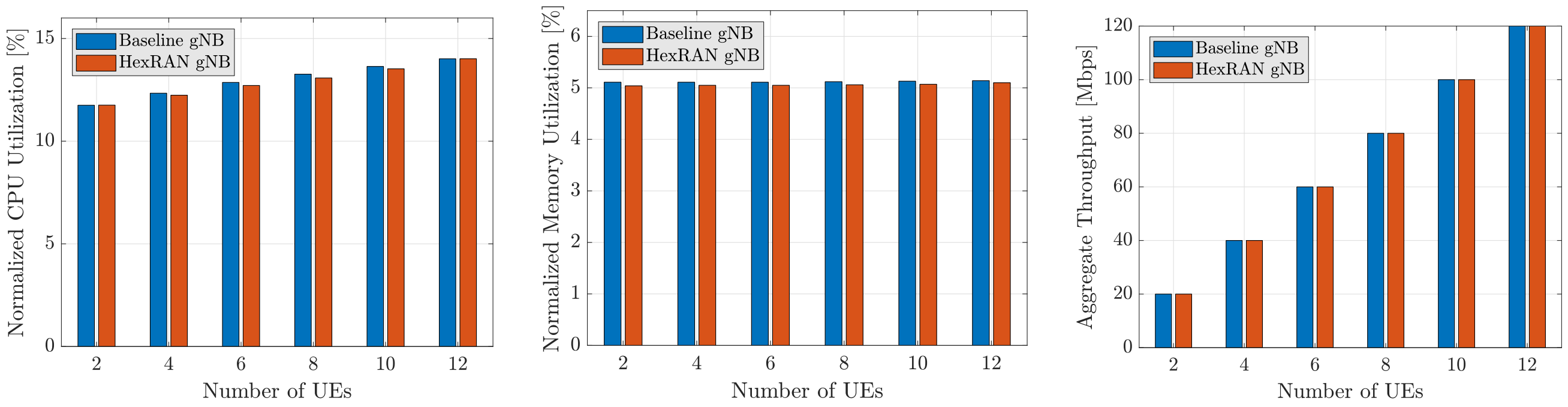

(a) CPU utilization comparison.

(b) Memory utilization comparison.

(c) Aggregate throughput comparison.

Fig. 12: Compute utilization and throughput comparison between HexRAN and the state-of-the-art.

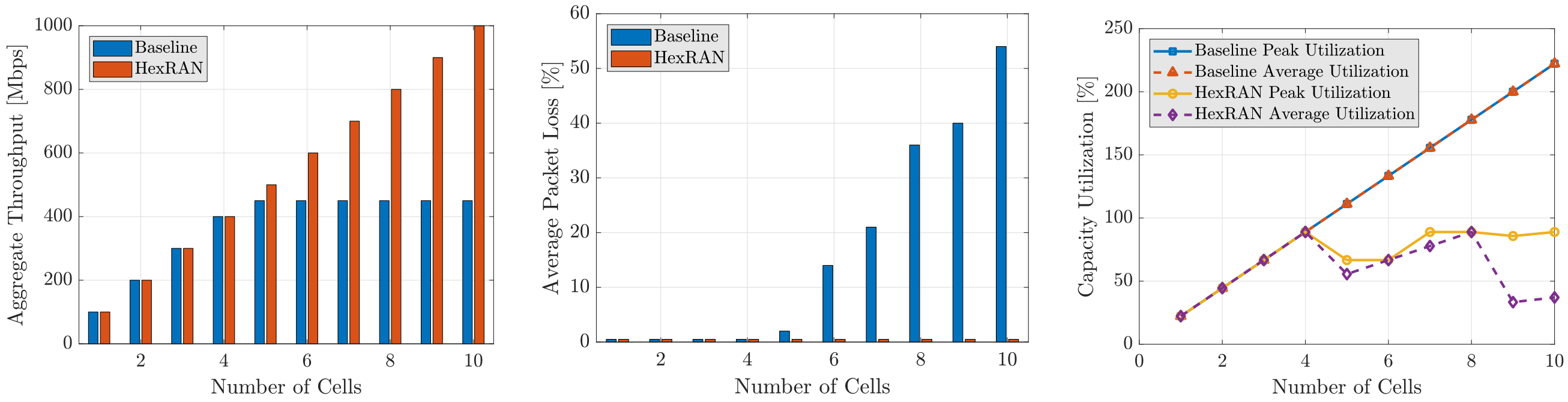

(a) Aggregate throughput comparison.

(b) Packet loss comparison.

(c) Capacity utilization comparison.

Fig. 13: Performance comparison between HexRAN and the disaggregated baseline.

120 Mbps for 12. Thus, it is apparent that HexRAN achieves performance parity with the baseline, thereby showcasing that the protocol decomposition offered by the HSS results in no performance penalty.

Having established that the HexRAN architecture does not introduce a performance overhead, we now showcase performance enhancements that are a direct outcome of the scalability inherent within the HSS. To that end, as part of the next experiment, we compare the performance of a distributed HexRAN deployment to that of a disaggregated baseline solution consisting of the CU and DU. In particular, the HexRAN deployment consists of six types of HUs– HU Type 1 consisting of low-PHY and high-PHY services, HU Type 2 which includes user scheduling services, HU Type 3 consisting of buffering and segmentation services, HU Type 4 offering user plane data encryption, HU Type 5 offering QoS enforcement services, and HU Type 6 offering cell, user, and bearer management services. HU Types 1 through 3 are deployed on the same node, while Types 4 and 5 are deployed on another node, with HU Type 6 being deployed on a third node. We note that each of the five HU types can be scaled independently with a new instance being deployed in the event that the capacity utilization on any given HU exceeds 90%.

Within this context, by leveraging the emulated radio environment, we benchmark HexRAN's performance against the baseline in terms of three metrics– the overall aggregate throughput achieved by the network, the average packet loss, and the per-NF capacity utilization. The experiment is conducted by increasing the number of cells in the network from 1 to 10, wherein, for HexRAN, the addition of each cell entails the deployment of an additional instance of HU Types 1 through 3, while for the baseline, each new cell requires the deployment of a new DU. Each cell carries 100 Mbps of downlink traffic, with the resulting performance comparison being showcased in Fig. 13. More specifically, from Fig. 13a, we note that as the number of cells in the network increases, the aggregate traffic also increases, with both HexRAN and the baseline offering identical performance up to four cells. However, upon the addition of a fifth cell, the baseline suffers a degradation in performance, while HexRAN continues to satisfy the increased traffic demand. Consequently, as the number of cells increase to 10, HexRAN seamlessly delivers the required 1 Gbps of traffic, while the baseline remains stagnant at 450 Mbps. This result follows from the fact that the modular nature of the HU and HSS allows for the instantiation of additional HU instances (of Types 4 and 5 in this case) to cater to the increase in traffic that occurs with the deployment of additional cells. The results presented in Fig. 13b reinforce this assertion with the baseline suffering rising levels of packet loss with an increase in the cell count, while HexRAN suffers from no packet loss.

HexRAN's inherent scalability advantage is further apparent in Fig. 13c, which benchmarks the peak and average capacity utilization levels across all NFs within the network. For HexRAN, each NF represents an HU instance (of Types 1 through 5), while for the baseline, each NF represents either a DU or a CU. We have previously noted that HexRAN automatically deploys a new HU instance if the utilization of a given HU type exceeds 90%, followed by load balancing across all instances of the same HU type. Consequently, as shown in the figure, the network-wide utilization across all HUs remains below 90% at all times, ensuring that HexRAN is able to maintain peak performance. On the other hand, the coarse-grained CU-DU approach of the baseline does not offer the same level of fine-grained scalablity, with the peak utilization hitting 220% of the available capacity, thereby resulting in the performance degradation observed in Figs. 13a and 13b.

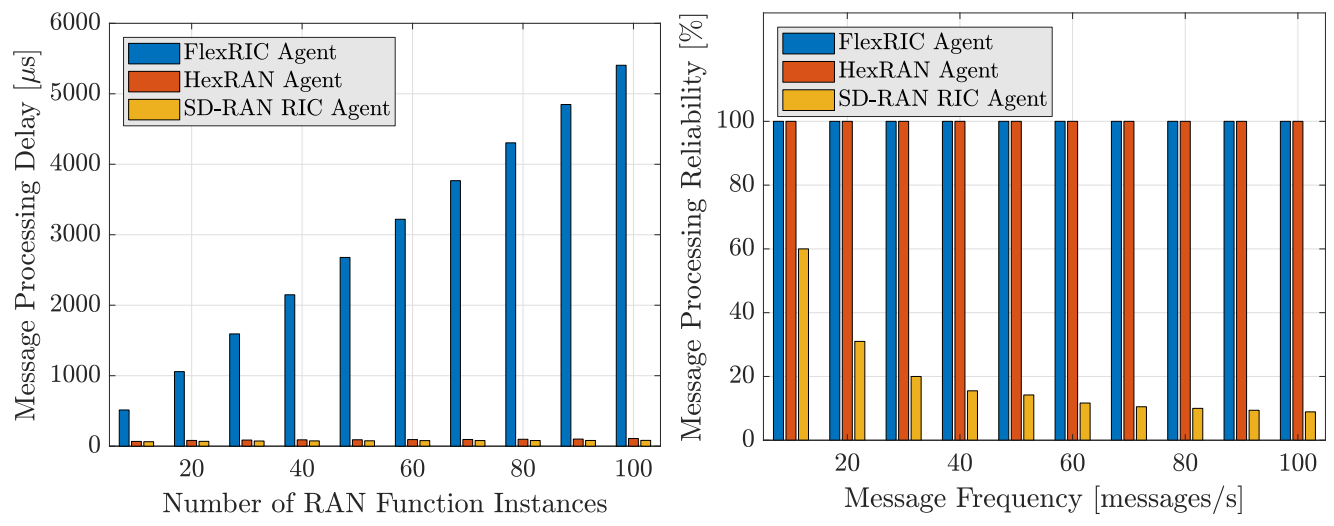

(a) Processing delay. (b) Processing reliability.

Fig. 14: HexRAN Agent performance comparison.

### D. Benchmarking the HexRAN Agent

Since the HexRAN Agent (HA) serves as the nerve-center for O-RAN within the base station, its performance has a tangible impact on all aspects of RAN operations ranging from reliable telemetry to high-performance assurance and control. While we have detailed the architectural novelties of the HA in §IV-C, a quantitative comparison with state-of-the-art solutions is equally imperative for establishing the performance advantage offered by HexRAN. To that end, in this section, we benchmark the HA against two open source O-RAN Agent implementations– the FlexRIC Agent [42][1] and the SD-RAN RIC Agent [43][2]. In the interest of a fair comparison, we implement the FS RAN function on both the FlexRIC and SD-RAN Agents, with all three platforms, i.e., HexRAN, OAI with the FlexRIC Agent, and SD-RAN, being deployed on the hardware radio environment.

We fist compare the message processing delay across all three agents as the number of RAN function instances increases from 10 to 100. For each instance that is added, the corresponding near-RT RIC sends a control request every 50 ms, and we measure the time elapsed between when a request message is received at the E2 interface and when the corresponding RAN function action execution API is called. The delay metric thus obtained represents the message processing delay showcased in Fig. 14a. From the figure we note that while the message processing delay for the FlexRIC Agent increases linearly, hitting 5500 $\mu$s for 100 instances, the delay values for both the HA and SD-RAN's RIC Agent remain relatively constant at around 100 $\mu$s. This result follows from the fact in the FlexRIC Agent, incoming messages have to wait for all prior messages to finish execution. As the number of RAN function instances increases, each arriving message ends up to having to wait for an increasingly longer duration, thereby impacting the system's scalability. On the other hand, both HexRAN and SD-RAN employ decoupled processing and are therefore not bound by execution delays.

Next, we compare message processing reliability across the three agents, wherein reliability represents the ratio of messages for which the corresponding control action is executed to the total number of control messages received by the RAN. To that end, we increase the message frequency from 10 to 100 messages per second, with the results thus obtained being showcased in Fig. 14b. From the figure we see that both the HA and FlexRIC Agent offer 100% reliability across all message frequencies, while SD-RAN's message processing reliability falls with an increase in the message frequency, reaching 9% for a message frequency of 100. This follows from the fact that the execution of slice-related control messages within SD-RAN is tied to the MAC scheduler, with only one control message being executed every frame, i.e., 10 ms. Furthermore, SD-RAN also lacks a message queuing mechanism, with pending control messages being overwritten repeatedly. Ultimately, as the message frequency increases, the reliability plummets. Thus, we can conclude that not only is the HexRAN Agent extremely feature rich, it also offers high performance characterized by low delays and robust reliability, qualities which are essential for next-generation RANs.

### E. Enabling RAN Slice Programmability

In §V we have detailed RAN slicing as a key use case for HexRAN with the FlexiSlice Framework (FSF) providing relevant plugins for the Programmable Mediation Layer (PML) and HexRAN Agent (HA) components with a view to enabling RAN slice programmability. In this section, we quantify the performance impact of the FSF through a series of experiments aimed at analyzing the throughput, latency, and resource utilization associated with slices in different states, as well as the relative performance of users having different bearer priorities.

We begin by deploying a HexRAN base station broadcasting a 40 MHz (i.e., 106 RBs) n78 carrier that can achieve a maximum downlink throughput of 130 Mbps. The gNB is configured with two RAN slices, 1 and 2, that are initially operating in the shared state, and are thus allocated an equal share of the radio resources, as shown in Fig. 15. Then, at $t$ = 20 s, in response to a change in the network policy, FS Configuration xApp on the near-RT RIC sends a control request that changes the configuration for Slice 2 to dedicated with 85 RBs causing its throughput to rise to 103 Mbps, at the expense of shared Slice 1. However, since both slices have sufficient resources to meet their respective target throughputs, congestion is not present, as evidenced by the relatively low round-trip time (RTT) values. Next, at $t$ = 40 s, all active users leave Slice 2 causing its traffic to drop to zero, at the same time, a second change in the network policy mandates a 130 Mbps downlink throughput for Slice 1. Since Slice 2

[1]Commit SHA-1 ID 2547045286adfa59927a375b535ac2c243f52588

[2]Commit SHA-1 ID 39f09cc369c3d726e02de720096c19ba5b72aee0

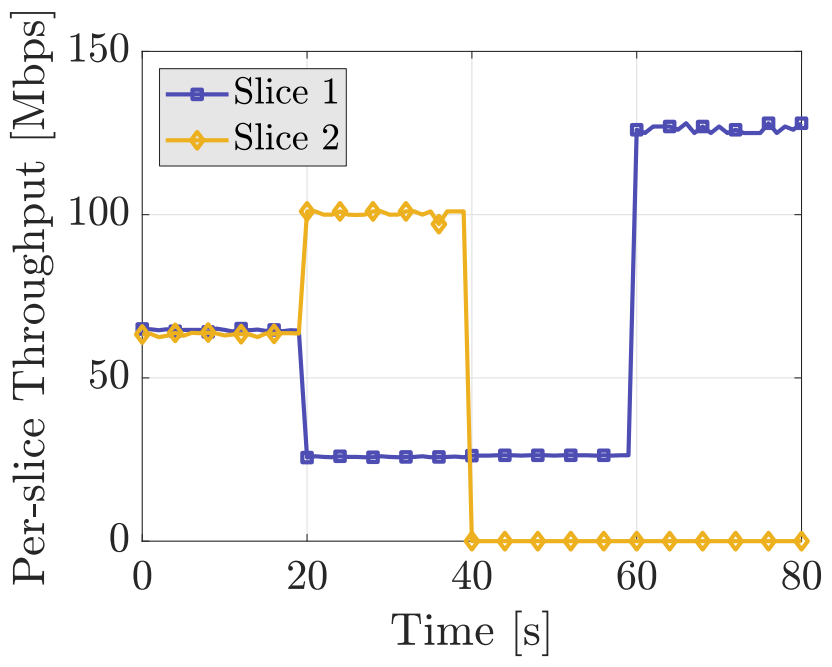

(a) Per-slice throughput.

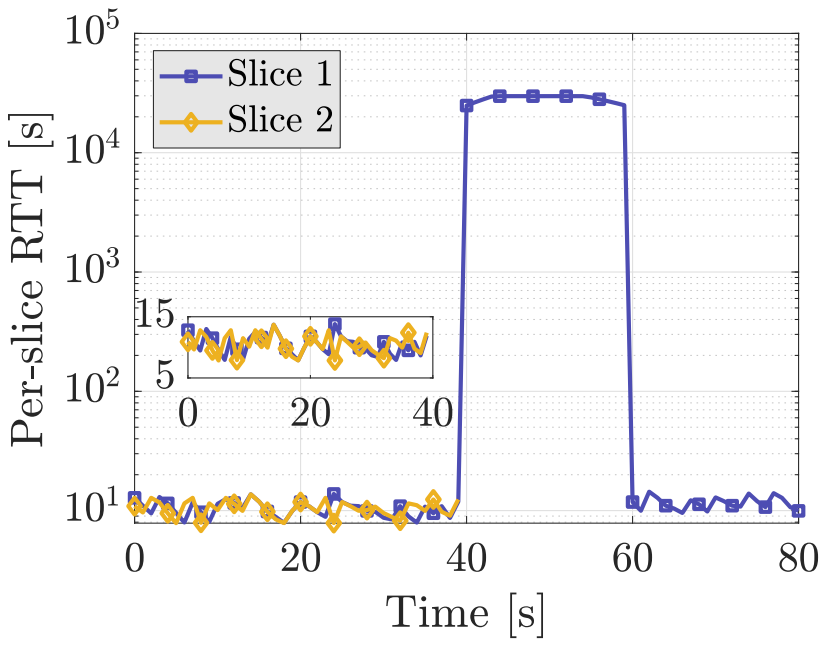

(b) Per-slice round-trip time.

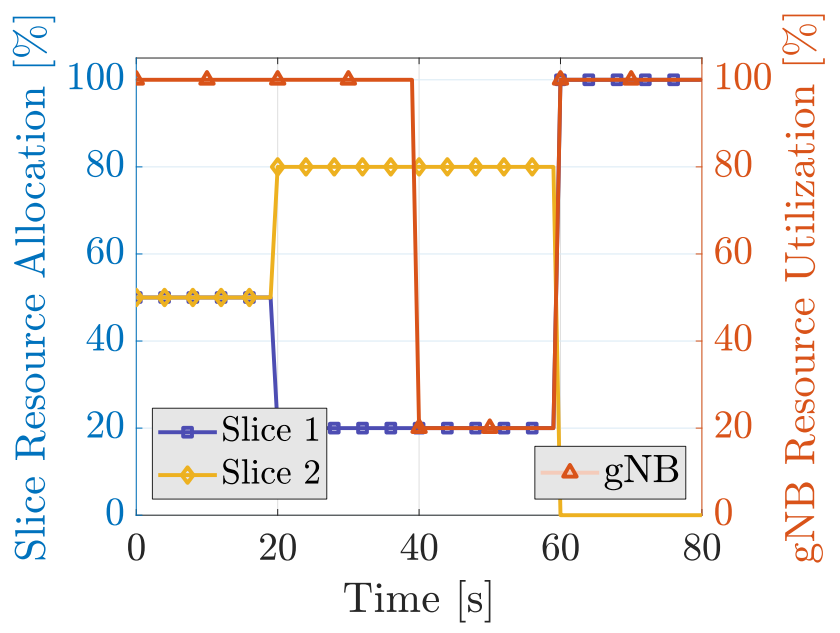

(c) Resource allocation and utilization.

Fig. 15: Performance impact of the FlexiSlice framework.

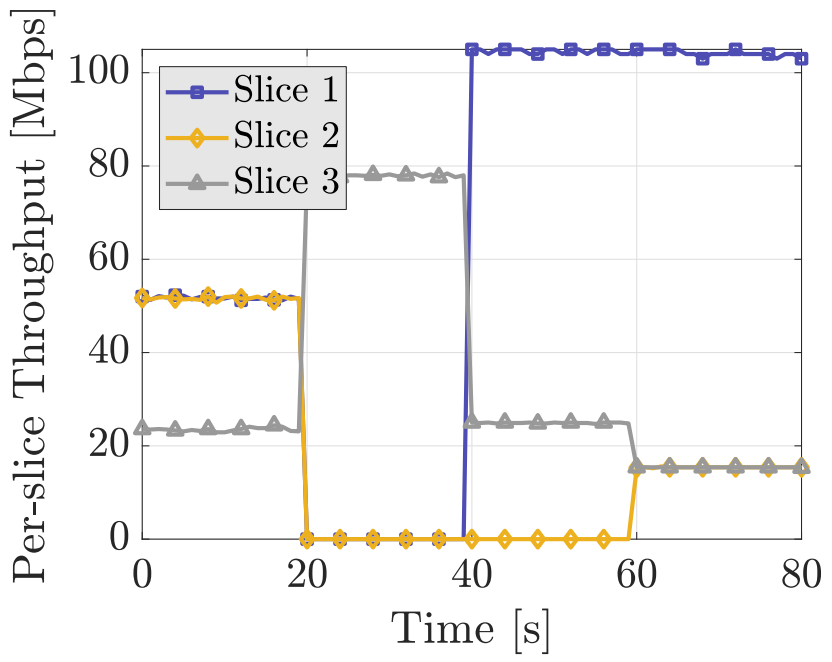

(a) Per-slice throughput.

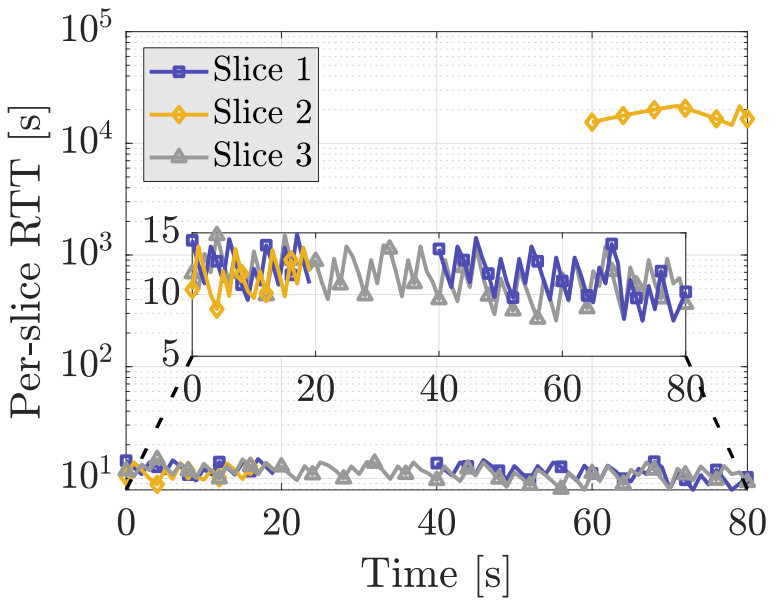

(b) Per-slice round-trip time.

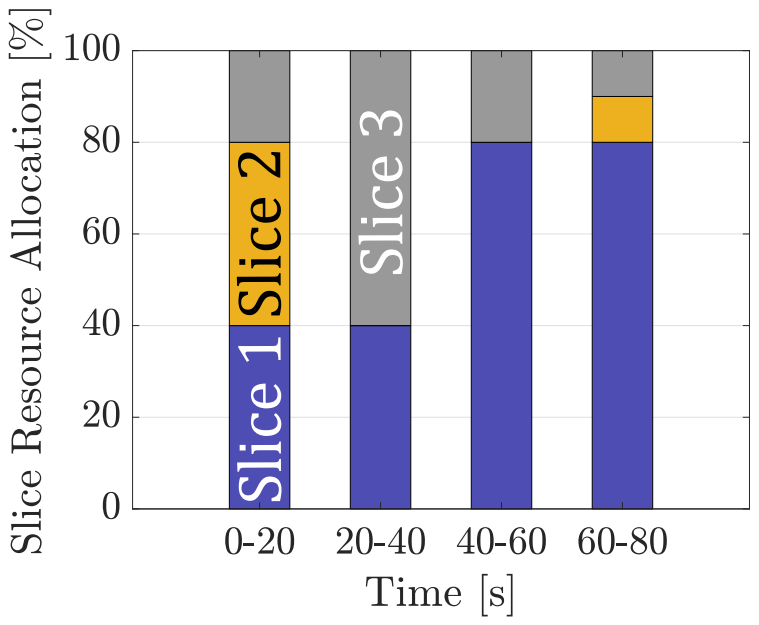

(c) Slice-level resource allocation.

Fig. 16: Comparing different slice states under dynamic network conditions.

is in the dedicated state, its unused resources do not revert to the shared pool.

Following from Fig. 15, we note that overall gNB resource utilization plummets to 20%, and Slice 1 is not able to increase its throughput beyond 27 Mbps, leading to congestion and a sharp spike in its RTT. To restore performance, at $t$ = 60 s, a second control request from the RIC changes the state of Slice 2 to prioritized. Now, all unused resources from Slice 2 go to the shared pool, and are thus available to Slice 1. Consequently, Slice 1 is able to achieve its target downlink throughput of 130 Mbps. In this manner, while the dedicated and prioritized states can be used to enforce both resource isolation as well as fine-grained performance requirements, the flexibility to switch between different states ensures that HexRAN is able to adapt to changing network conditions. However, the wastage of unused resources associated with dedicated slices might lead one to question the utility of the dedicated state. However, as the next experiment will show, dedicated slices do have a critical role to play in the network.

For the next experiment, we deploy three slices on the gNB– Slice 1 in the dedicated state, Slice 2 in the prioritized state, and Slice 3 in the shared state. We note that since the hybrid slice represents an amalgamation of the dedicated, prioritized, and shared states, we do not deploy a fourth hybrid slice. As shown in Fig. 16, from $t$ = 0 – 19 s, Slices 1 and 2 are each assigned 40% of the available resources, with the shared pool containing the remaining 20% for Slice 3. With the throughput requirements for each slice being in line with their respective resource assignments, the network does not suffer from congestion. Then, at $t$ = 20 s, all active users leave Slices 1 and 2. Slice 1 remains in the dedicated state, while Slice 2 gives up its prioritized resource assignment and switches to the idle state. Thus, Slice 3 now has access to 60% of the network's resources, and its downlink throughput rises to 78 Mbps. Next, at $t$ = 40 s, Slice 1 gains active users, and at the same time, owing to a change in network policy, the FS Configuration xApp assigns 80% of the available resources to Slice 1. Predictably, from the figure, we note a rise in throughput for Slice 1 and a fall for Slice 3.

Then, at $t$ = 60 s, Slice 3 too gains active users and transitions from the idle state. Since Slice 3 was prioritized previously, it now moves to the shared state, albeit with a target throughput of 52 Mbps as before. However, at this point, both Slices 2 and 3 are sharing the same shared pool having only 20% of the total resources. With a 10% resource assignment, Slice 2 is not able to meet its target throughput, resulting in a congestion-induced spike in its RTTs. These results underscore the importance of the dedicated state. The ability to switch to the dedicated mode is ideal for critical use cases such as public safety networks during emergencies, thereby guaranteeing a minimum level of performance that is isolated from changing network conditions. Once the emergency scenario is over, these slices can be switched back to the prioritized mode in the interest of resource efficiency. This experiment also illustrates the performance isolation offered by HexRAN, with Slices 1 and 3 being unaffected by the congestion in Slice 2.

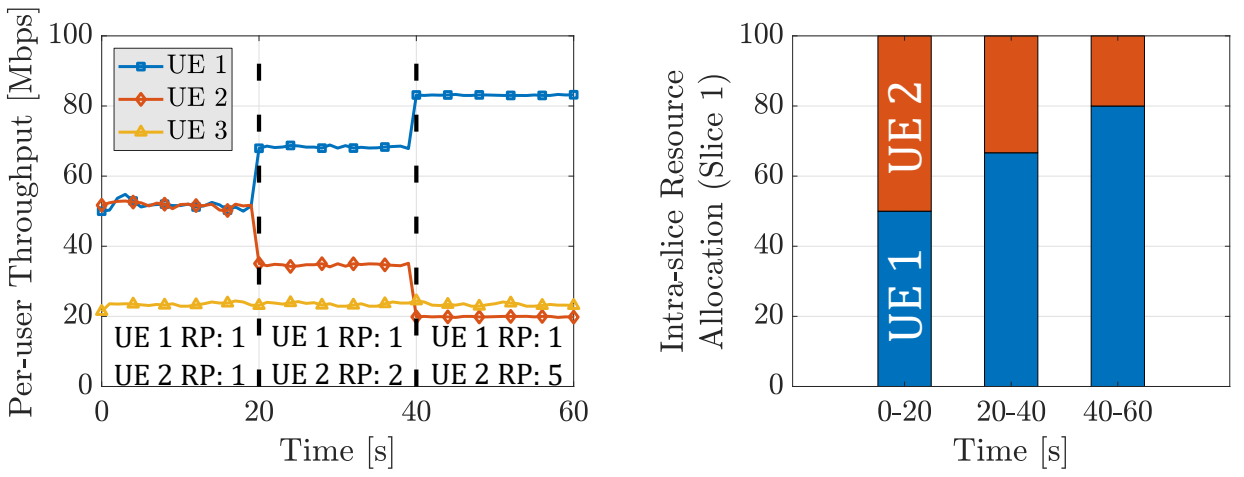

(a) Per-user throughput. (b) Slice 1 allocation.

Fig. 17: Bearer priority-based user prioritization.

As noted in §V, the FS framework incorporates a UE-level configuration parameter that allows for the dynamic adjustment of bearer priorities, providing a RAN-level mechanism to differentiate between users that would otherwise have the exact same QoS configuration as set by the core. We demonstrate this capability in the following experiment by considering a scenario with two slices– Slice 1 in the dedicated state and Slice 2 in the shared state. Slice 1 contains two UEs 1 and 2, while Slice 2 contains UE 3. All three users are configured with the same 5QI (5G QoS Identifier) value, 9. At $t = 0$, the FS Configuration xApp assigns 85 RBs to Slice 1, and this assignment remains fixed throughout the experiment. Our focus here is on the two UEs in Slice 1, both of which are initially configured with a bearer priority (BP) value 1, and thus share Slice 1's resources equally, as shown in Fig. 17.

As shown in the figure, at $t = 20$ s, the xApp changes the BP for UE 2 to 2. Within the context of this experiment, an increase in the bearer priority metric is interpreted as a decrease in user priority, consequently, UE 2 is allocated a lower share of the resources, and its throughput falls. A second change in UE 2's BP to 5 at $t = 40$ s, leads to further deprioritization, and a 80 – 20 resource split in favor of UE 1, which increases its throughput to 83 Mbps. In this manner, HexRAN allows for fine-grained control over the users' QoS performance, which is particularly useful in scenarios such as infrastructure sharing, wherein tenants may not have control over classical QoS parameters such as the 5QI.

## VII. Conclusion

In this paper, we have presented HexRAN– a programmable approach to O-RAN base station design. HexRAN introduces several novel features relating to RAN disaggregation and composability, 3GPP and O-RAN protocol integration and programmability, robust RAN-RIC communications, and customizable RAN slicing. Furthermore, we have demonstrated vertically integrated RAN programmability enabled by HexRAN through the FlexiSlice framework. We have also implemented a prototype of the HexRAN architecture, deployed it on a lab-scale over-the-air testbed, and conducted a comprehensive performance evaluation addressing the key themes of system scalability, agent performance, and RAN slice programmability. The results thus obtained have validated that not only is HexRAN the most feature-complete RAN architecture, it also offers a significant performance advantage over the prior art.